\documentclass[aps,pre,superscriptaddress,twocolumn,balancelastpage]{revtex4-1}

\usepackage{times}
\usepackage{amsmath,amssymb}
\usepackage{graphicx}

\usepackage{verbatim}
\usepackage{color}

\usepackage{placeins}     

\usepackage{physics}

\newcommand{\ue}{\text{e}}
\newcommand{\ui}{\text{i}}

\newcommand{\RhoOmega}{\rho_{V_{12}}}
\newcommand{\RhoCN}{\rho_{\text{CN}}}

\newcommand{\RhoNNSDistRegu}{\widetilde{P}}
\newcommand{\RhoNNSUndist}{P_0}

\newcommand{\sRTwo}{s^{\text{R}_2}}
\newcommand{\sCN}{s^{\text{CN}}}
\newcommand{\sNNSDist}{s}
\newcommand{\sNNSUndist}{s^{(0)}}
\newcommand{\sMeanRho}{\overline{\mathbf{s}}}

\newcommand{\jkNext}{j^1k^1}
\newcommand{\jkCN}{j^2k^2}

\makeatletter
\let\Hy@backout\@gobble
\makeatother

\newcommand{\ud}{\text{d}}

\newcommand{\Marcenko}{Mar{\v c}enko}
\newcommand{\cU}{{\cal U}}
\newcommand{\cE}{{\cal E}}

\newcommand{\kt}{\rangle}
\newcommand{\br}{\langle}

\newcommand{\rhoA}{\rho_1}
\newcommand{\rhoB}{\rho_2}

\DeclareMathOperator\erfc{erfc}

\begin{document}

\title{Entanglement in coupled kicked tops with chaotic dynamics}

\author{Tabea Herrmann}
\affiliation{Technische Universit\"at Dresden,
 Institut f\"ur Theoretische Physik and Center for Dynamics,
 01062 Dresden, Germany}

\author{Maximilian F.~I. Kieler}
\affiliation{Technische Universit\"at Dresden,
 Institut f\"ur Theoretische Physik and Center for Dynamics,
 01062 Dresden, Germany}

\author{Felix Fritzsch}
\affiliation{Technische Universit\"at Dresden,
 Institut f\"ur Theoretische Physik and Center for Dynamics,
 01062 Dresden, Germany}

\author{Arnd B\"acker}
\affiliation{Technische Universit\"at Dresden,
 Institut f\"ur Theoretische Physik and Center for Dynamics,
 01062 Dresden, Germany}
\affiliation{Max-Planck-Institut f\"ur Physik komplexer Systeme,
N\"othnitzer Stra\ss e 38, 01187 Dresden, Germany}

\date{\today}

\begin{abstract}
The entanglement of eigenstates in two coupled, classically chaotic
kicked tops is studied in dependence of their interaction strength.
The transition from the non-interacting and unentangled system towards full
random matrix behavior is governed by a universal scaling
parameter.
Using suitable random matrix transition ensembles we express this transition
parameter as a function of the subsystem sizes and the coupling strength for
both unitary and orthogonal symmetry classes.
The universality is confirmed for the level spacing statistics
of the coupled kicked tops
and a perturbative description is in good agreement with numerical results.
The statistics of Schmidt eigenvalues and entanglement
entropies of eigenstates is found to follow a universal scaling as well.
Remarkably this is not only the case for large subsystems of equal size
but also if one of them is much smaller.
For the entanglement entropies a perturbative description is
obtained, which can be extended to large couplings
and provides very good agreement with numerical results.
Furthermore, the transition of the statistics of the entanglement
spectrum towards the random matrix limit is demonstrated
for different ratios of the subsystem sizes.
\end{abstract}

\maketitle

\section{Introduction}

Entanglement is one of the key features of quantum theory
and, besides of being of fundamental conceptual interest \cite{EinPodRos1935},
has nowadays many applications
ranging from quantum communication, quantum cryptography, to
quantum computing \cite{HorHorHorHor2009,NieChu2010, DebLinFigLanWriMon2016}.
It also plays an important role
in characterizing phases of quantum many-body systems
\cite{AmiFazOstVed2008,PolSenSilVen2011,BorIzrSanZel2016,AleKafPolRig2016,ParPotVas2017,AbaPap2017,LuiBar2017b}.
A fundamental condition for entanglement is a system consisting of multiple
interacting subsystems.
The simplest model to investigate entanglement properties of
such systems are bipartite systems, which consist of two
subsystems coupled by some interaction.
One of the central questions for such bipartite systems
concerns the possible amount of entanglement, quantified e.g.\
by the von Neumann entropy, R\'enyi entropies,
Havrda-Charv{\'a}t-Tsallis (HCT) entropies, or the Schmidt
eigenvalues~\cite{Neu1932,Ren1961wcrossref,HavCha1967,Tsa1988,BenZyc2006}.
This question concerns both the entanglement generated
in the time-evolution of initially un-entangled states
and the entanglement of eigenstates of the full system.
A common scenario is that the subsystems are ``quantum-chaotic''
in the sense that their spectral statistics and eigenstate
statistics are well-described by random matrix theory.
Such systems could have a classical limit with chaotic dynamics,
while in the context of many-body systems
a classical limit not necessarily exists.

If the subsystems are strongly coupled, their bipartite entanglement
can be obtained
from a random matrix description.
This implies that the statistics of Schmidt eigenvalues is given
by the \Marcenko-Pastur distribution~\cite{MarPas1967,SomZyc2004}
and leads to predictions for the average values of the purity
(or linear entropy) \cite{Lub1978} and von Neumann entropy
\cite{Pag1993,Sen1996}.
These results apply for example to quantum systems
with classically chaotic dynamics, as shown for
coupled standard maps \cite{Lak2001}
or coupled kicked tops \cite{MilSar1999,BanLak2002,BanLak2004},
and to chaotic states in many body systems, see e.g.\
Refs.~\cite{KhlKru2014,BeuAndHaq2015,PonPapHuvAba2015,GarGro2018,
  BeuBaeMoeHaq2018,HacVidRigBia2019}.

If the subsystems are not strongly interacting, the amount of eigenstate
entanglement is reduced.
For bipartite systems with broken time-reversal symmetry
this has been intensively studied in the
last few years \cite{SriTomLakKetBae2016,LakSriKetBaeTom2016,TomLakSriBae2018}:
a universal transition from
unentangled to entangled states was found to be determined by a single
transition parameter depending only on the system sizes and the interaction
strength. Furthermore a random matrix transition ensemble was
introduced which allows to describe the universal features of
entanglement and spectral statistics.
Moreover a perturbation theoretical description
for spectral statistics (consecutive level spacing distribution)
and different measures of entanglement has been
developed. For the entanglement entropies a recursively applied embedded
perturbation theory describes the whole transition
towards maximal entanglement \cite{LakSriKetBaeTom2016,TomLakSriBae2018}.
Recently, a perturbative description of the
time-dependence of entanglement entropies for initial product
eigenstates was obtained which leads
to a universal prediction after an appropriate rescaling of
time \cite{PulLakSriBaeTom2019:p}.

In this paper we study eigenstate entanglement in bipartite systems
with and without time reversal symmetry and different types of couplings
between the subsystems, based on techniques developed in
Refs.~\cite{SriTomLakKetBae2016,LakSriKetBaeTom2016,TomLakSriBae2018}.
To illustrate the analytical results we use
a pair of coupled kicked tops as a system with time reversal invariance
for both equal and different subsystem dimensions.
The kicked top model was set up to study the influence
of classical regular and chaotic behavior on
quantum mechanical properties \cite{HaaKusSch1987, AriEvaSar1992, Haa2010}.
Coupled kicked tops have been introduced
to investigate the time evolution of entanglement \cite{MilSar1999a},
and since then explored in much detail, see
e.g.\ Refs.~\cite{MilSar1999a, MilSar1999, BanLak2002, TanFujMiy2002,
  FujMiyTan2003, ZniPro2003, BanLak2004, DemKus2004, TraMadDeu2008,
  KubAdaTod2008, KubAdaTod2013, PucPawZyc2016, KumSamAna2017, AdaKubTod2019}.
Kicked tops are of particular interest, as they can
also be accomplished experimentally
\cite{ChaSmiAndGhoJes2009,NeiEtAl2016,MeiAngAnGad2019,KriAnjBhoMah2019,
  MunPogJesDeu2019:p}
and realizing coupled kicked tops might therefore be feasible
in the future.
This would also provide a possibility
to probe entanglement in a coupled many-body system
as the total spin of each subsystem
can be considered as the sum of spin-1/2 qubits
\cite{Mil1999:p,WanGhoSanHu2004,DogMadLak2019,SieOlsElbHeyHauHaaZol2019}.
We study the eigenstate entanglement for coupled kicked tops
when both show fully chaotic behavior in the classical
limit. For this we derive the transition parameter $\Lambda$ for the general
case of systems with time reversal invariance
and specifically for the random matrix transition ensemble
with random diagonal coupling.
To account for the specific interaction of the coupled kicked tops
it turned out to be necessary to introduce
a random matrix transition ensemble
with random product phases for the coupling.
Furthermore we develop a perturbation theory of the level spacing
statistics for same subsystem dimensions and find a prediction for the
uncoupled situation for different subsystem dimensions.
To describe the entanglement of the coupled kicked tops
in dependence on the transition parameter,
we use a perturbative description for the
first two Schmidt eigenvalues and for the entanglement entropies.
Applying the recursive embedding of the regularized
perturbation theory, following Ref.~\cite{TomLakSriBae2018},
leads to a description of the complete transition.
Good agreement with numerical calculations for same as well as
for different subsystem dimension is found.
In addition we show that the distribution of the Schmidt eigenvalues
approaches  the \Marcenko-Pastur distribution
for large transition parameters, even though quite slowly.

The paper is organized as follows:
In Sec.~\ref{subsec:bipartite_system} we introduce bipartite systems
and their time evolution operator
and in Sec.~\ref{subsec:universal_transition_para}
define the transition parameter for which a general expression
is obtained if the individual subsystems
can be described by random matrix theory.
Section~\ref{sec:RMTE} discusses
different random matrix transition ensembles
with their transition parameters and statistical properties.
In Sec.~\ref{subsec:coupled_kicked_tops} we introduce
the coupled kicked tops and
the transition parameter for this system.
In Sec.~\ref{sec:level_spacing_stat} the level spacing statistics
is studied and a perturbative description is derived for the case of
equal subsystem dimension and also
the case of different subsystem dimensions is considered.
Using the level spacing distribution we
demonstrate the universality of the transition parameter. In
Sec.~\ref{sec:entanglement} we study the entanglement and its perturbative
description for coupled kicked tops. For this we introduce in
Sec.~\ref{subsec:schmidt_eigenvalues_entropies} the Schmidt eigenvalues and the
entanglement entropies as measures for the entanglement in bipartite
systems. In Sec.~\ref{subsec:schmidt_eigenvalues} we present perturbation theory
results for the first two Schmidt eigenvalues. This perturbation theory
is extended in Sec.~\ref{subsec:entropies} to the entanglement entropies and
the recursively embedded perturbation theory is employed
to describe the whole transition. In
Sec.~\ref{subsec:entropies_diff_dim} we discuss the applicability
of this theory to
the case of different subsystem dimensions and in
Sec.~\ref{subsec:stat_schmidt_eigenvalues} the
full statistics of the Schmidt eigenvalues is considered.
Finally, a summary and outlook is given in Sec.~\ref{sec:summary}.

\section{Bipartite systems}\label{subsec:bipartite_system}

We consider a class of interacting bipartite systems
in which the time evolution is
given by a unitary Floquet operator, i.e.\
the propagator over one period of the driving, as
\begin{equation}
    \cU = U_{12}(\varepsilon) (U_1\otimes U_2).
    \label{eq:U_bipartite_system}
\end{equation}
Here $U_1$ and $U_2$ are unitary operators on
Hilbert spaces of dimension $N_1$ and $N_2$, respectively,
and $U_{12}(\varepsilon)$
acts on the tensor product space of dimension $N_1 N_2$
and provides a coupling between the two subsystems.
The coupling is assumed
to fulfill $U_{12}(0) = \text{Id}$,
i.e.\ there is no interaction
between the subsystems for $\varepsilon=0$.
With increasing
$\varepsilon$ the interaction increases and the operator $U_{12}(\varepsilon)$
is assumed to be entangling \cite{ZanZalFao2000,PalLak2018}.
Its eigenvalue problem is given by
\begin{equation}
   \cU |\psi_n \rangle = \exp(\ui \varphi_n) |\psi_n \rangle
\end{equation}
with eigenstates  $|\psi_n\rangle$
and corresponding eigenvalues $\exp(\ui \varphi_n)$, which
lie on the unit circle due to the unitarity of $\cU$,
so that the eigenphases $\varphi_n \in [0, 2\pi[$.
We aim to characterize the statistics
of eigenphases and eigenstates in dependence on the strength
$\varepsilon$ of the coupling
and the Hilbert space dimensions $N_1$ and $N_2$.

\subsection{Universal transition parameter}
           \label{subsec:universal_transition_para}

In various cases
the statistical properties of the bipartite system~\eqref{eq:U_bipartite_system}
turn out to be governed by a single scaling parameter $\Lambda$
\cite{PanMeh1983,FreKotPanTom1988, SriTomLakKetBae2016, LakSriKetBaeTom2016,
TomLakSriBae2018,PulLakSriBaeTom2019:p}.
This universal transition parameter is given by
\begin{equation} \label{eq:lambda_definition}
  \Lambda = \frac{v^2}{D^2},
\end{equation}
where $v^2$ is the mean square off-diagonal matrix element
of $U_{12}(\varepsilon)$ in the basis in which $U_1 \otimes U_2$ is diagonal
and $D=\frac{2\pi}{N_1 N_2}$ is the mean level spacing
of the full system.

For systems in which the non-interacting subsystems
$U_1$ and $U_2$ can be modeled by
random unitary matrices chosen from an appropriate ensemble
the transition parameter depends on the coupling
$U_{12}$ only.
Specifically, in the absence of anti-unitary symmetries
$U_1$ and $U_2$ are chosen from the
circular unitary ensemble (CUE) while the
circular
orthogonal ensemble (COE) applies in the presence of
an anti-unitary symmetry (e.g.\ time-reversal) \cite{Meh2004}.

The ensemble average for the COE leads to
\begin{equation}     \label{eq:lambda_coe}
\begin{split}
  \Lambda_{\text{COE}}  =
 & \frac{N_1N_2}{4\pi^2(N_1N_2- 1)(N_1+2)(N_2+2)}  \\
 & \times \Bigl(N_1N_2\bigl(N_1N_2+2(N_1+N_2)\bigl) \Bigr. \\
 & \quad \Bigl. - 2 ||U_{12}^{(1)}||^2
  - 2 ||U_{12}^{(2)}||^2 - |\text{tr}(U_{12})|^2\Bigl)\, ,
\end{split}
\end{equation}
which is derived in App.~\ref{app:COE-transition-parameter}.
Here $U_{12}^{(1)}$ and $U_{12}^{(2)}$
are diagonal matrices with entries
$(U_{12}^{(1)})_{jj}=\sum_{k}(U_{12})_{jk,jk}$,
and $(U_{12}^{(2)})_{kk}=\sum_{j}(U_{12})_{jk,jk}$
as partially traced interaction operators,
which are in general not unitary, and
$\|X\|^2=\Trace(XX^{\dagger})$ is the Hilbert-Schmidt norm
\cite{SriTomLakKetBae2016, TomLakSriBae2018}.

If the subsystems have equal dimension, $N=N_1=N_2$,
Eq.~\eqref{eq:lambda_coe} simplifies to
\begin{equation}       \label{eq:lambda_coe_N}
  \begin{split}
  \Lambda_{\text{COE}} =
           &  \frac{N^4}{4\pi^2 (N^2 -1)(N+2)^2}
            \biggl( N^2 + 4N \\
           & \qquad - 2 \;\biggl|\biggl|\frac{U_{12}^{(1)}}{N}\biggl|\biggl|^2
    - 2 \;\biggl|\biggl|\frac{U_{12}^{(2)}}{N}\biggl|\biggl|^2  -
      \; \biggl|\frac{\text{tr}(U_{12})}{N}\biggl|^2 \biggl) \,.
\end{split}
\end{equation}

For the CUE one gets
\begin{equation}\label{eq:lambda_cue}
\begin{split}
  \Lambda_{\text{CUE}} =
      & \frac{N_1N_2}{4\pi^2(N_1N_2- 1)(N_1+1)(N_2+1)} \\
      & \qquad\times \Bigl(  N_1N_2\bigl(N_1N_2+ (N_1+N_2)\bigl) \\
      & \qquad \quad - ||U_{12}^{(1)}||^2   -  ||U_{12}^{(2)}||^2
                     - |\text{tr}(U_{12})|^2\Bigl),
\end{split}
\end{equation}
which is derived in App.~\ref{app:CUE-transition-parameter}.
For $N=N_1=N_2$ this simplifies to
\begin{equation}\label{eq:lambda_cue_N}
\begin{split}
  \Lambda_{\text{CUE}} =
       & \frac{N^4}{4\pi^2 (N^2 -1)(N+1)^2} \biggl( N^2 + 2N \\
       & \qquad
               -  \;\biggl|\biggl|\frac{U_{12}^{(1)}}{N}\biggl|\biggl|^2
               -  \;\biggl|\biggl|\frac{U_{12}^{(2)}}{N}\biggl|\biggl|^2
               -  \; \biggl|\frac{\text{tr}(U_{12})}{N}\biggl|^2 \biggl) .
\end{split}
\end{equation}
Note, that Eq.~\eqref{eq:lambda_cue_N} differs slightly
from the result obtained in Refs.~\cite{SriTomLakKetBae2016, TomLakSriBae2018},
but agrees in leading order for example
with the results of the random matrix transition ensemble, see
Eq.~\eqref{eq:lambda-CUE-a-la-TP} below.

The above expressions for the transition parameter
show that to obtain the same value of $\Lambda$
for different Hilbert space dimensions $N_1$ and $N_2$
the coupling strength $\varepsilon$ has
to be adjusted accordingly.
The explicit dependence on $\varepsilon$ is governed by
the specific form of the coupling.

\subsection{Random matrix transition ensembles}
\label{sec:RMTE}

\subsubsection{General random matrix transition ensemble}

To define explicit random matrix models to describe the statistical
properties of eigenvalues and eigenstates and the transition parameter
of bipartite systems of the form \eqref{eq:U_bipartite_system}
one has to prescribe the statistical properties of the coupling.
The general form of the random matrix transition ensemble is
\begin{equation}  \label{eq:U_rmt-general}
   \cU_{\text{RMT}}(\varepsilon)
     = U_{12}(\varepsilon) (U_1^{\text{RMT}}\otimes U_2^{\text{RMT}}),
\end{equation}
where $U_1^{\text{RMT}}$ and $U_2^{\text{RMT}}$ are random matrices, e.g.\
from the COE or the CUE.
The coupling is written as
\begin{equation} \label{eq:U-12-via-V-12}
  U_{12}(\varepsilon)=\exp(\ui \varepsilon V_{12})
\end{equation}
and a rather general modeling is given by a diagonal matrix
\begin{equation}
  (V_{12})_{jk, j^{\prime}k^{\prime}}=2\pi \xi(j, k)
   \delta_{jj^{\prime}}\delta_{kk^{\prime}},
\end{equation}
with $j,j'=1, ..., N_1$ and $k,k'=1, ..., N_2$.
The phase $\xi(j, k)$ is assumed to be random
following some prescribed distribution.

\subsubsection{Random matrix transition ensemble}
\label{subsec:RMTE}

In Ref.~\cite{SriTomLakKetBae2016} the random matrix transition
ensemble was introduced for which the coupling is given by
\begin{equation}
   \label{eq:rmt-TE-diagonal-coupling}
  (V_{12})_{jk, j^{\prime}k^{\prime}}=2\pi\xi_{jk}
   \delta_{jj^{\prime}}\delta_{kk^{\prime}},
\end{equation}
where $\xi_{jk}$ are i.i.d.\ distributed uniformly on $[-1/2, 1/2]$.
The limiting case of strong coupling has been studied in
Ref.~\cite{LakPucZyc2014}, where the
entangling power of ${\cal U}_{\text{CUE}}(\varepsilon=1)$ was derived
analytically.

Using the general result \eqref{eq:lambda_coe} for the COE case gives,
see App.~\ref{app:COE-CUE-ensemble-transition-parameter},
\begin{equation}
\begin{split}
\Lambda_{\text{COE}} = & \dfrac{N_1^2 N_2^2}{4 \pi^2(N_1 + 2)(N_2 + 2)}  \\
    & \times \frac{(N_1 + 2) (N_2+2) - 9}{N_1 N_2 - 1}
 \left[1-\dfrac{\sin^2 (\pi \varepsilon)}{\pi^2\varepsilon^2} \right],
\end{split}
\label{eq:lambda-COE-TP}
\end{equation}
and for the CUE,
\begin{equation}
\begin{split}
\Lambda_{\text{CUE}} = & \dfrac{N_1^2 N_2^2}{4 \pi^2(N_1 + 1)(N_2 + 1)}  \\
   & \times \frac{(N_1 + 1) (N_2+1) - 4}{N_1 N_2 - 1}
 \left[1-\dfrac{\sin^2 (\pi \varepsilon)}{\pi^2\varepsilon^2} \right] .
\end{split}
\label{eq:lambda-CUE-a-la-TP}
\end{equation}

In the definition of the transition
parameter~\eqref{eq:lambda_definition} the off-diagonal
elements of $U_{12}$ appear in the numerator.
Thus, when applying the perturbation theory below
to describe the spectral statistics and the entanglement
in dependence on $\Lambda$, the distribution
of the matrix elements
\begin{equation} \label{eq:omega-jk-def}
  \omega_{jk} = \frac{1}{\tilde{v}^2} |\bra{j^{\prime}k^{\prime}}V_{12}\ket{jk}|^2,
\end{equation}
in the eigenbasis $\ket{jk}$ of the uncoupled
system plays an important role.
Here $\tilde{v}$ is the mean square off-diagonal element
of $V_{12}$ in this basis such that $\omega_{jk}$ has unit mean.
For small $\varepsilon$ one has
$U_{12}(\varepsilon)=\exp(\ui \varepsilon V_{12})
\approx \text{Id} + \ui \varepsilon V_{12}$
and thus $v^2 = \varepsilon^2 \tilde{v}^2$.
As there are no correlations between the matrix elements,
the coupling \eqref{eq:rmt-TE-diagonal-coupling}
leads for the COE case to $\omega_{jk}$ following the
Porter-Thomas distribution \cite{PorTho1956}
\begin{equation} \label{eq:porter-thomas}
  \RhoOmega(\omega) = \frac{1}{\sqrt{2\pi\omega}} \exp(-\omega/2).
\end{equation}
For the CUE transition ensemble one gets the exponential
\begin{equation} \label{eq:exponential}
  \RhoOmega(\omega) =  \exp(-\omega).
\end{equation}

\subsubsection{Random matrix transition ensemble with product phases}
\label{subsec:RMTE-KT-coupling}

The coupling \eqref{eq:rmt-TE-diagonal-coupling}
provides the simplest possible form
and  leads to a good description of spectral statistics and entanglement
in a wide class of systems
\cite{SriTomLakKetBae2016, LakSriKetBaeTom2016, TomLakSriBae2018,
PulLakSriBaeTom2019:p}.
However one may have other types of interactions
leading to different expressions for the transition
parameter and the statistical properties.
A physically relevant case occurs when the matrix $V_{12}$ is the tensor
product of matrices acting on the individual subsystems Hilbert spaces.
In this case the phases can be described by
a product of random individual phases,
\begin{equation} \label{eq:rmt-TE-KT-diagonal-coupling}
  (V_{12})_{jk, j^{\prime}k^{\prime}}=
     2\pi\xi_{j} \tilde{\xi}_k  \delta_{jj^{\prime}}\delta_{kk^{\prime}},
\end{equation}
where $\xi_{j}$ and $\tilde{\xi}_k$ are i.i.d.\ distributed
uniformly on $[-1/2, 1/2]$.
Using the general result \eqref{eq:lambda_coe} for the COE case gives
for small $\varepsilon$,
see App.~\ref{app:COE-CUE-ensemble-transition-parameter-product-phases},
\begin{equation}
\begin{split}
  \Lambda_{\text{COE}} \simeq &
  \frac{\varepsilon^2}{144} \frac{(N_1N_2)^2\left((N_1+1)(N_2+1)-9\right)}
                                 {(N_1N_2-1)(N_1+2)(N_2+2)},
\end{split}
\label{eq:lambda-COE-TE-KT-coupling}
\end{equation}
and for the CUE
\begin{equation}
\begin{split}
  \Lambda_{\text{CUE}} \simeq
    \frac{\varepsilon^2}{144} \frac{(N_1N_2)^2\left((N_1+1)(N_2+1)-4\right)}
                                   {(N_1N_2-1)(N_1+1)(N_2+1)}.
\end{split}
\label{eq:lambda-CUE-TE-KT-coupling}
\end{equation}
The full expressions, valid for larger $\varepsilon$ as well,
are also given in
App.~\ref{app:COE-CUE-ensemble-transition-parameter-product-phases}.

Moreover, due to the product structure of the phases, the distribution of the
matrix elements $\omega_{jk}$ is given by
\begin{equation}
  \RhoOmega(\omega) = \frac{1}{\pi \sqrt{w}} K_0(\sqrt{w})\;,
  \label{eq:RMTE-KT-coupling-K0-distrib}
\end{equation}
see App.~\ref{ap:matrix_element_dist_KT}, where $K_0$ is the modified
Bessel function of the second kind \cite[Eq.~10.25.3]{DLMFCurrent}.
For the CUE case one gets
\begin{align}
  \RhoOmega(\omega) = 2 K_0(2\sqrt{\omega})\;.
  \label{eq:RMTE-KT-coupling-K0-distrib-CUE}
\end{align}

\subsection{Example: Coupled kicked tops}
\label{subsec:coupled_kicked_tops}

As specific example of an interacting bipartite system
we consider a pair of coupled
time-periodically kicked tops,
which have been studied in much detail
in particular with respect to entanglement generation,
see e.g.\ Refs.~\cite{MilSar1999a, MilSar1999, BanLak2002, TanFujMiy2002,
  FujMiyTan2003, ZniPro2003, BanLak2004, DemKus2004, TraMadDeu2008,
  KubAdaTod2008, KubAdaTod2013, PucPawZyc2016, KumSamAna2017}.
The dynamics is described by the Hamiltonian \cite{MilSar1999a,BanLak2002}
\begin{align}
    H(t) = H_1(t) + H_2(t) + H_{12}(t),
\end{align}
where
\begin{align}
    H_\ell(t) =&\, \frac{\pi}{2} J_{y_\ell}
               + \frac{k_\ell}{2j_\ell} (J_{z_\ell} + \alpha_\ell)^2
                 \sum_{n=-\infty}^{\infty} \delta(t-n), \\
    H_{12} =\,& \varepsilon \frac{1}{\sqrt{j_1j_2}} J_{z_1}J_{z_2}
    \sum_{n=-\infty}^{\infty} \delta(t-n)
    \label{eq:hamilton_kickedtop_4D}\;.
\end{align}
Here $j_\ell$
is the total angular momentum of the $\ell$-th spin ($\ell=1,2$), and
$J_{y_\ell}$ und $J_{z_\ell}$ are the components of the angular momentum
operator.
Although the following equally applies to half integer spins we
for simplicity restrict the discussion to integer $j_\ell$.
The parameters $k_\ell$ are the individual kicking strengths
of the two tops and $\varepsilon$
determines the coupling strength between the two tops.
For $\varepsilon=0$ the two subsystems are uncoupled.
The Hilbert spaces of the uncoupled spins
have dimension $N_1 = 2 j_1 + 1$ and $N_2 = 2 j_2 + 1$, respectively.
The real parameters $\alpha_\ell$
are additional phases which allow to break the
parity symmetry \cite{BanLak2002}.

The Floquet operator
for the coupled tops is given by \cite{BanLak2002}
\begin{equation} \label{eq:U_kickedtop_4D}
    \cU = U_{12}(\varepsilon) (U_1 \otimes U_2)\,,
\end{equation}
where
\begin{equation} \label{eq:U_kickedtop_2D}
  U_\ell = \exp\left(-\frac{\ui k_\ell}{2j_\ell}
          \left(J_{z_\ell}+\alpha_\ell\right)^2 \right)
  \exp\left(-\frac{\ui \pi}{2}J_{y_\ell}\right) \,,
\end{equation}
and the coupling reads
\begin{align} \label{eq:U_12_kickedtop_4D}
  U_{12}(\varepsilon)=\exp(\ui \varepsilon V_{12})
  \quad \text{with}\ \ V_{12}=\frac{1}{\sqrt{j_1 j_2}} J_{z_1} J_{z_2} .
\end{align}
The order of the operators is such that
we consider the free evolution first and then apply the kicks.

In the following we use $k_1=12.0$ and $k_2=15.0$
for which the classical dynamics
corresponding to each top in the uncoupled case
(numerically) shows chaotic motion
with no visible regular structures.
As phases we use $\alpha_1 = 0.35$ and $\alpha_2 = 0.4$
so that there is only time-reversal symmetry \cite{BanLak2002}.
Therefore the Floquet operators $U_\ell$ for the individual spins
and their spectral statistics can be modeled by the COE.

To compute the transition parameter for the coupled kicked tops
we replace $U_\ell$ by independent COE matrices to
use the general COE result \eqref{eq:lambda_coe}
and compute the specific expressions for the
coupling \eqref{eq:U_12_kickedtop_4D}, see
Eqs.~\eqref{eq:U_12_trace_kickedtop_approx}--\eqref{eq:U_12_norm_partial_trace_2_kickedtop_approx}
in App.~\ref{app:KT-transition-parameter}.
In the numerical computations these expressions
are used to determine $\Lambda$ in dependence on $\varepsilon$ and $j_\ell$.
For large $N_1$, $N_2$ and small $\varepsilon$ one gets
\begin{equation}
  \Lambda \approx
      \frac{1}{144 \pi^2}\varepsilon^2 N_1 N_2 [N_1N_2 + 2(N_1+N_2)].
\end{equation}
As discussed before in Sec.~\ref{subsec:universal_transition_para},
this expression explicitly shows
that to get the same $\Lambda$ for different $j_1$ and $j_2$
one has to adapt the coupling accordingly.

\section{Level spacing statistics}
\label{sec:level_spacing_stat}

\subsection{Level spacing statistics for equal dimensions}
\label{subsec:level_spacing_stat}

To demonstrate that the transition parameter
indeed leads to a universal description for the coupled kicked tops,
we first consider the distribution of consecutive level spacings
for equal Hilbert space dimensions $N=N_1=N_2$.
The distribution $P(s)$ of the (re-scaled) consecutive
level spacings $s_n = \tfrac{1}{D}(\varphi_{n+1} - \varphi_n)$,
where $D$ is the mean level spacing,
depends on the strength of the coupling between the subsystems:
For strong coupling
$P(s)$ should follow the results of the COE \cite{BohGiaSch1984}
which is well-described by the Wigner distribution
\begin{equation} \label{eq:Ps-GOE}
  P_{\text{COE}}(s) \approx
      \frac{\pi}{2} s \exp\left(-\frac{\pi}{4}s^2\right) .
\end{equation}
For the uncoupled case, even though the individual
subsystems show COE statistics, the resulting
spacing distribution of the full bipartite system for large $N_1$ and $N_2$
approaches the exponential
\begin{equation} \label{eq:Ps-Poissonian}
  P_{\text{Poisson}}(s) = \exp(-s) .
\end{equation}
The reason for this is that the eigenphases of the full system
\eqref{eq:U_bipartite_system} are an independent superposition
$\varphi_{jk} = \theta_j^{(1)} + \theta_k^{(2)} \text{mod } 2\pi$
of the phases $\theta_j^{(1)}$ and $\theta_k^{(2)}$
of the individual subsystems, respectively,
where $j=1, ..., N_1$ and $k=1, ..., N_2$.
Note that for tensor products of CUE matrices of equal dimension
it has been proven in Ref.~\cite{TkoSmaKusZeiZyc2012}
that the spectral statistics become Poissonian.

\begin{figure*}
  \includegraphics{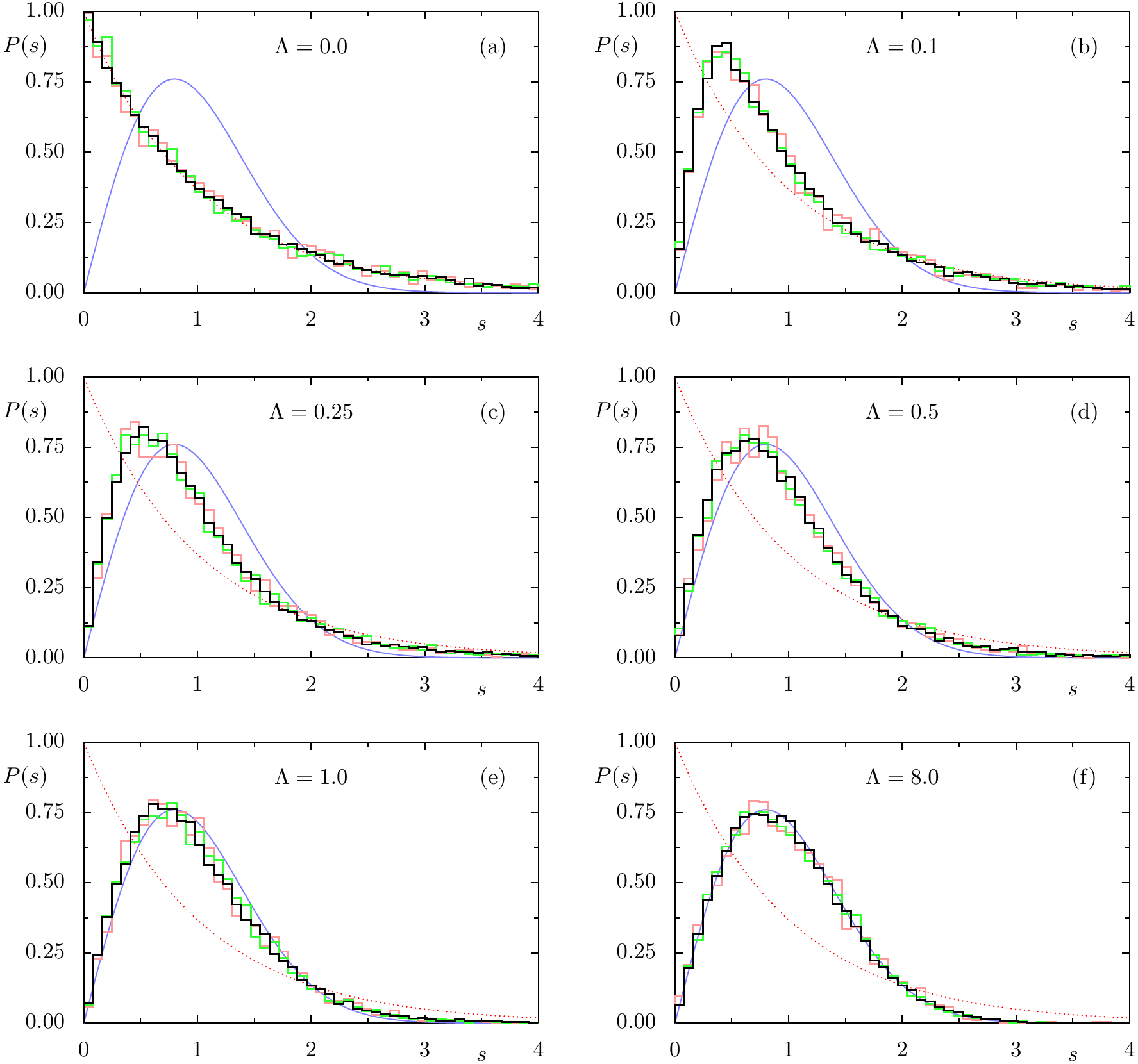}

  \caption{Transition of the level spacing distribution
    for the coupled kicked tops for
    (a) $\Lambda=0.0$,
    (b) $\Lambda=0.1$,
    (c) $\Lambda=0.25$,
    (d) $\Lambda=0.5$,
    (e) $\Lambda=1.0$, and
    (f) $\Lambda=8.0$.
    The exponential \eqref{eq:Ps-Poissonian},
    is shown as red dotted curve
    and the COE result \eqref{eq:Ps-GOE},
    as blue solid curve.
    In each case the histograms for
    $j_1=j_2=30$ (very light red),
    $j_1=j_2=50$ (light green), and
    $j_1=j_2=70$ (black) are shown.
    The other parameters are
     $k_1=12.0$, $k_2=15.0$, $\alpha_1 = 0.35$,
     and $\alpha_2 = 0.4$.}
  \label{fig:coupled_kickedtop_spacings}
\end{figure*}

\begin{figure}
  \includegraphics{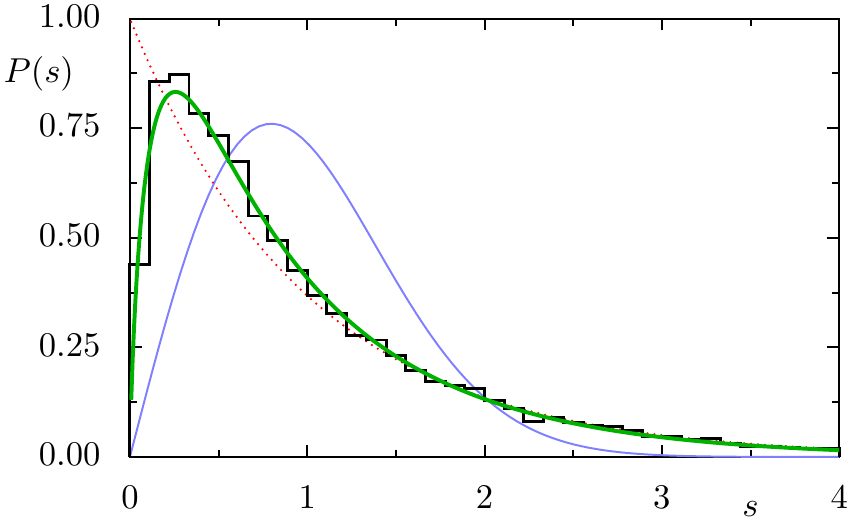}

  \caption{Perturbative prediction of the level spacing distribution
    for the coupled kicked tops at small
    $\Lambda = 0.02$.
    Shown is the histogram for $j_1=j_2=70$,
    $k_1=12.0$, $k_2=15.0$, $\alpha_1 = 0.35$, and
    $\alpha_2 = 0.4$.
    The thick green curve shows the result of the perturbation theory
    \eqref{eq:Ps-perturbative-prediction}.
    For comparison the result \eqref{eq:Ps-GOE}
    for the COE (solid blue curve),
    and the exponential \eqref{eq:Ps-Poissonian}
    (red dotted curve) are shown.}
\label{fig:coupled_kickedtop_spacing_same_lambda}
\end{figure}

Figure~\ref{fig:coupled_kickedtop_spacings}
shows the level spacing distribution $P(s)$
for different $j_1=j_2=30, 50, 70$ in dependence on $\Lambda$.
At $\Lambda=0$ one obtains good agreement with the exponential
\eqref{eq:Ps-Poissonian}.
Once $\Lambda>0$ there is an instantaneous
change to level-repulsion, i.e.\ $P(0)=0$,
as illustrated in Fig.~\ref{fig:coupled_kickedtop_spacings}(b)
for $\Lambda=0.1$.
Increasing $\Lambda$ further one gets
closer to the result for the COE, Eq.~\eqref{eq:Ps-GOE}.
While the initial change of the distribution is rather
rapid in $\Lambda$, this slows down at around $\Lambda=1.0$
and the COE statistics is well fulfilled at $\Lambda=8.0$,
see Fig.~\ref{fig:coupled_kickedtop_spacings}(f).
Interestingly, this happens significantly later than
in case of the CUE transition ensemble
and the coupled kicked rotors on the torus
where good agreement is found at $\Lambda=1.0$ \cite{SriTomLakKetBae2016}.
Numerical results for the transition ensemble \eqref{eq:U_rmt-general}
with COE matrices and interaction \eqref{eq:rmt-TE-diagonal-coupling}
show the same slower approach to the COE limit ~\eqref{eq:Ps-GOE}.
Thus this is an inherent feature of the considered COE case
and not specific to the coupling \eqref{eq:U_12_kickedtop_4D}
of the kicked tops.

The sequence of plots in Fig.~\ref{fig:coupled_kickedtop_spacings}
confirms that $\Lambda$ provides the universal scaling
parameter: For the same $\Lambda$, but
different $j_1=j_2$ and corresponding $\varepsilon$,
determined implicitly using \eqref{eq:lambda_coe} and
Eqs.~\eqref{eq:U_12_trace_kickedtop}--\eqref{eq:U_12_norm_partial_trace_2_kickedtop},
the histograms nicely fall on top of each other.

In order to derive a perturbative expression for $P(s)$ we
closely follow the derivation in the CUE case given in
Ref.~\cite{TomLakSriBae2018} and adapt this to the COE
transition ensemble.
We choose the specific interaction for the coupled kicked tops
given by Eq.~\eqref{eq:U_12_kickedtop_4D}
for which $V_{12}$ is a tensor product of spin operators
acting on the individual subsystems.
Thus we model its statistics by a random interaction of product form
defined in Eq.~\eqref{eq:rmt-TE-KT-diagonal-coupling}.
The starting point is the perturbative expansion of the eigenphases
$\varphi_{jk}$ of $\cU$ according to \cite{TomLakSriBae2018}
\begin{align}
  \varphi_{jk} = \theta_{jk} + \varepsilon \bra{jk}V_{12}\ket{jk}
              + \varepsilon^2 \sum_{j^{\prime}k^{\prime}\neq jk}
              \frac{|\bra{j^{\prime}k^{\prime}} V_{12}\ket{jk}|^2}
              {\theta_{jk}-\theta_{j^{\prime}k^{\prime}}}.
\end{align}
Here $\theta_{jk}$
are the eigenphases of the unperturbed system, i.e.\ for $\varepsilon=0$,
corresponding to the eigenstates
$\ket{jk}$, with $j=1, ..., N_1$ and $k=1, ...., N_2$,
which form a basis of the full Hilbert space.
To compute the distribution of the normalized
consecutive level spacings
\begin{align}
\sNNSDist_{jk} = \frac{\varphi_{jk}-\varphi_{\jkNext}}{D},
\end{align}
where $\ket{\jkNext}$ is the eigenstate for which
$\theta_{\jkNext}$ is the consecutive eigenphase of $\theta_{jk}$ in the
unperturbed system
we take the average over the random matrix
transition ensemble with product phases,
see Sec.~\ref{subsec:RMTE-KT-coupling}.
Doing so not all terms in the perturbative
expression contribute to the  consecutive level spacing.
In particular, the first order correction merely shifts the whole spectrum
leaving all level spacings unchanged.
Moreover, as the
eigenphases $\theta_{jk}$ are uniformly distributed in $\left[0, 2\pi\right[$
only two second order terms $\sim \varepsilon^2$ contribute
to the difference $\varphi_{jk}-\varphi_{\jkNext}$ upon averaging.
For simplicity we only keep these non-vanishing terms which
leads to the level spacing
\cite{SriTomLakKetBae2016}
\begin{align}
  \sNNSDist_{jk} = \frac{\theta_{jk}-\theta_{\jkNext}}{D}
                +2\frac{\varepsilon^2}{D}\frac{|\bra{jk}V_{12}\ket{\jkNext}|^2}
                  {\theta_{jk}-\theta_{\jkNext}}\;.
\end{align}

The averaging procedure is then performed by
replacing $|\bra{jk}V_{12}\ket{\jkNext}|^2$
by $\tilde{v}^2 \omega_{jk}$.
Further $(\theta_{jk}-\theta_{\jkNext})/D$
is substituted by the spacing $\sNNSUndist_{jk}$ in the unperturbed system.
Noting that $v^2 = \varepsilon^2 \tilde{v}^2$ is the mean squared off-diagonal
element of the full perturbation $\varepsilon V_{12}$ gives in lowest order
of $\varepsilon$
\begin{align}
  \sNNSDist_{jk} =  \sNNSUndist_{jk}
                  + 2\Lambda\frac{\omega_{jk}}{\sNNSUndist_{jk}}
  \label{eq:levelspacing_perturbation}
\end{align}
with $\Lambda$ as defined in Eq.~\eqref{eq:lambda_definition}.

In order to compute the statistics of the level spacings $\sNNSDist$ we average
over both random variables $\sNNSUndist_{jk}$ and $\omega_{jk}$.
The distributions $\RhoNNSUndist(s_0)$ of $\sNNSUndist_{jk}$ is the Poisson
distribution~\eqref{eq:Ps-GOE} and the distribution
$\rho_{V_{12}}(\omega)$ of $\omega_{jk}$ is given by
Eq.~\eqref{eq:RMTE-KT-coupling-K0-distrib}.
The singular behavior of
Eq.~\eqref{eq:levelspacing_perturbation} at $\sNNSUndist_{jk} \to 0$
is dealt with by regularization
\cite{Tom1986, FreKotPanTom1988,SriTomLakKetBae2016},
i.e.\ the replacement
\begin{align}
  s_0  + 2\Lambda \frac{\omega}{s_0 }
      \rightarrow \sqrt{s_0^2 + 4\Lambda\omega}\;.
\end{align}
This follows from degenerate perturbation theory
and correctly captures the repulsion of nearly degenerate levels.
Since this also provides the correct asymptotic behavior,
this replacement can be done in the entire range of integration.
The distribution of level spacing results in
\begin{align}
  \RhoNNSDistRegu(\sNNSDist) =& \int_0^{\infty} \ud s_0 \int_0^{\infty}
                              \ud \omega\; \RhoOmega(\omega)\RhoNNSUndist(s_0)
                              \nonumber \\
                              & \times \delta\left(\sNNSDist - \sqrt{{s_0}^2
                              + 4\Lambda \omega}\right).
\label{eq:levelspacing_integral}
\end{align}
Note, that due the regularization procedure
$\RhoNNSDistRegu(\sNNSDist)$ does not have the required unit mean.
This condition can be restored using a rescaling of $s$ by
\begin{align}
\sMeanRho = \int_0^{\infty} \ud s\; s \, \RhoNNSDistRegu(s),
\end{align}
such that the final result for the level spacing distribution reads
\begin{align} \label{eq:Ps-perturbative-prediction}
  P(s) &= \sMeanRho \;\RhoNNSDistRegu(\sMeanRho s).
\end{align}

The resulting prediction as well as the level spacing distribution for the
coupled kicked tops is shown in
Fig.~\ref{fig:coupled_kickedtop_spacing_same_lambda} for $\Lambda=0.02$.
Here, we  evaluate the integrals in Eq.~\eqref{eq:levelspacing_integral}
numerically.
This figure demonstrates that there is a good agreement
between the perturbation theory result and the data for the coupled kicked tops.
Extending the prediction beyond the perturbative regime
is an interesting open problem.
Note that for obtaining the perturbative result
\eqref{eq:Ps-perturbative-prediction}
the specific form of the coupling
\eqref{eq:U_12_kickedtop_4D} has been used.
Using the random matrix transition ensemble
with coupling given by Eq.~\eqref{eq:rmt-TE-diagonal-coupling}
would lead to a prediction having its maximum further to the right.

\subsection{Level spacing statistics for different dimensions}
\label{subsec:level_spacing_stat_diff_dim}

The results shown in
Fig.~\ref{fig:coupled_kickedtop_spacings} and
Fig.~\ref{fig:coupled_kickedtop_spacing_same_lambda}
confirm that the transition parameter provides a
universal scaling in the case of equal Hilbert space dimensions $N_1=N_2$.
Moreover, the above derivation allows to treat
different dimensional subsystems, i.e.
$N_1 \neq N_2$, as well.
In particular the case of $N_1 \ll N_2$
is of relevance as it corresponds to
one system with a small Hilbert space coupled
to a system  with chaotic dynamics  and a much larger Hilbert space
which could be considered as representing a heat-bath.

\begin{figure}
  \includegraphics{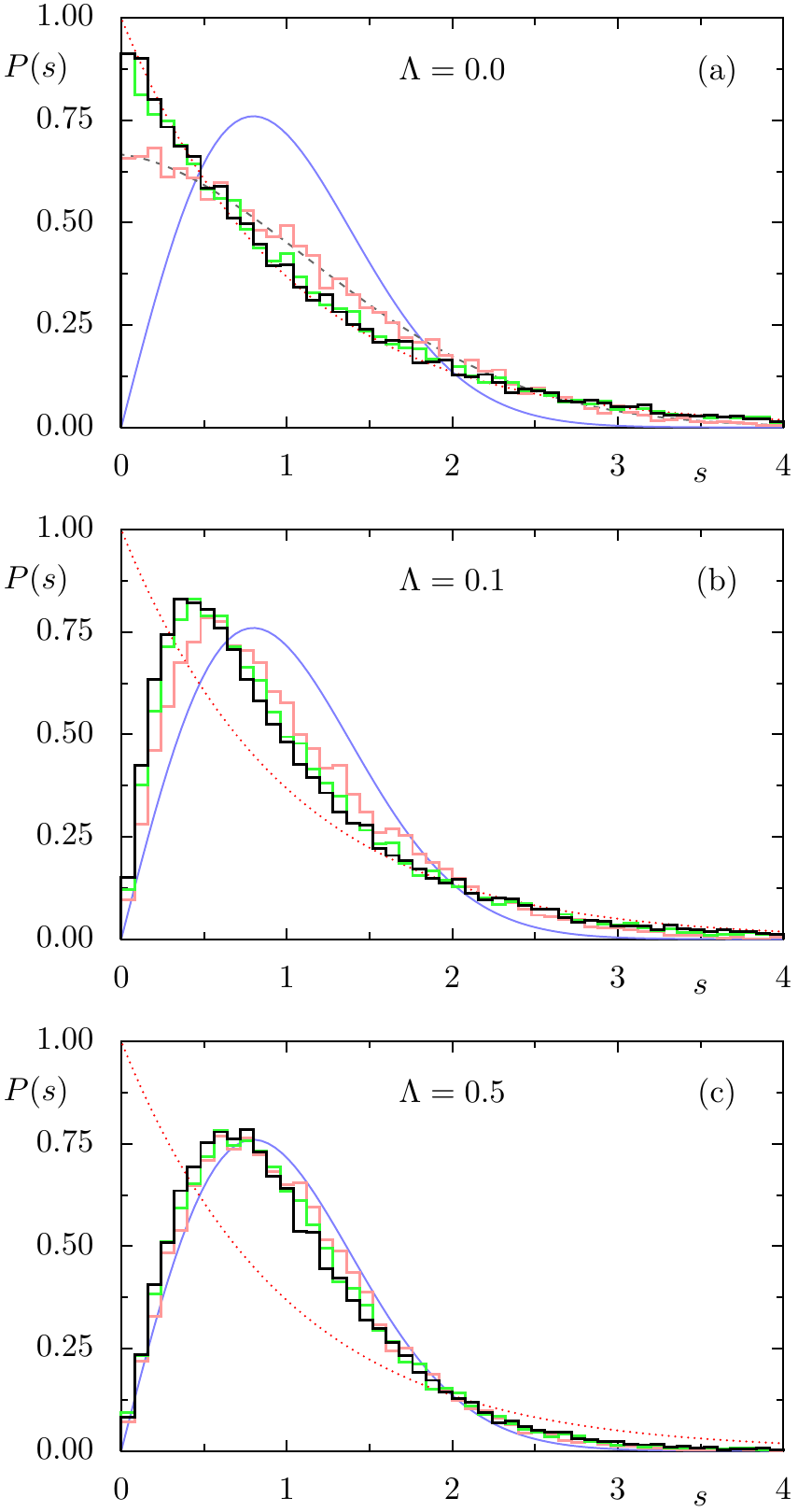}

  \caption{Level spacing statistics
           for different dimensions of the two coupled kicked tops
           for (a) $\Lambda=0.0$,
           (b) $\Lambda=0.1$, and (c) $\Lambda=0.5$.
           The histograms are for
           $(j_1, j_2) = (1, 3000)$, very light red,
           $(j_1, j_2)= (5, 1000)$, light green,
           and $(j_1, j_2)= (10, 500)$, black.
           The gray dashed curve in (a) shows the
           $N_1$COE statistics \eqref{eq:Ps-NCOE} for $N_1=3$.
           For comparison the result \eqref{eq:Ps-GOE}
           for the COE (solid blue curve),
           and the exponential \eqref{eq:Ps-Poissonian}
           (red dotted curve) are shown.
           The other parameters are
           $k_1=12.0$, $k_2=15.0$, $\alpha_1 = 0.35$, and $\alpha_2 = 0.4$.}
  \label{fig:Ps-different-dim}
\end{figure}

However, already for the uncoupled case, $\Lambda=0$, the
level spacing statistics $P(s)$
shows clear deviations from
the exponential behavior \eqref{eq:Ps-Poissonian},
see Fig.~\ref{fig:Ps-different-dim}(a).
This can be explained by the form
$\varphi_{jk} =
\theta_j^{(1)} + \theta_k^{(2)} \text{mod } 2\pi$
of the eigenphases of the uncoupled system.
If $N_1 \ll N_2$, the statistics of the second subsystem
is well described by the COE, so that the full spectrum can
be considered as a superposition of $N_1$ independent COE ensembles
of size $N_2$.
In this case the consecutive level spacing distribution
is given by the $N_1$COE statistics \cite{Abu2009},
\begin{equation} \label{eq:Ps-NCOE}
P_{N_1\text{COE}}(s) = t_1^{N_1-2} t_2 \left[ \left( 1 - \frac{1}{N_1} \right) t_2
	+ \frac{\pi s}{2 N_1^2} t_1 \right],
\end{equation}
where
\begin{equation}
   t_1 = \erfc \left( \frac{\sqrt{\pi} s}{2 N_1} \right)
   \quad \text{ and } \quad
   t_2 = \exp\left(-\frac{\pi s^2}{4 N_1^2}\right),
\end{equation}
using the complementary error function.
In Fig.~\ref{fig:Ps-different-dim}(a) we observe very good agreement
of Eq.~\eqref{eq:Ps-NCOE} with the numerical result
for the small dimension $N_1 = 3$, corresponding to $j_1=1$.
Furthermore we emphasise that Eq.~\eqref{eq:Ps-NCOE} converges
for $N_1 \to \infty$ to the
Poisson distribution \eqref{eq:Ps-Poissonian}.
This is already rather well achieved for $N_1 = 11$, corresponding to $j_1=5$,
which is also shown in Fig.~\ref{fig:Ps-different-dim}(a).

Thus the transition of the consecutive level
spacing distribution cannot be universal
if one subsystem dimension is small.
There are significant differences for different $j_1$
which are also present when the coupling is increased.
For strong coupling these differences disappear
as the two subsystem merge into one large system, whose
statistics becomes independent of the ratio of the subsystems sizes,
and is given by that of the COE.
We remark that the perturbation theory requires that both
subsystem dimensions are large. Here we considered the extreme case that one
dimension is small and thus it comes as no surprise that universality may
fail.

\section{Entanglement}\label{sec:entanglement}

\subsection{Schmidt eigenvalues and entanglement entropies}
\label{subsec:schmidt_eigenvalues_entropies}

For the bipartite system \eqref{eq:U_bipartite_system}
with tunable interaction the eigenstates of $\cU(0)$, i.e.\ the uncoupled case,
are simply product states of the individual subsystems
and therefore not entangled.
However with increasing interaction $\varepsilon$
they can no longer be written as product states, i.e.\
they become entangled.
To characterize the amount of entanglement between the
two subsystems, there exist different quantitative measures,
like the von Neumann entropy, R\'enyi entropies or
the Havrda-Charv{\'a}t-Tsallis
entropies~\cite{HavCha1967,Tsa1988,BenZyc2006}.
These measures are all based on the eigenvalues
of the reduced density matrices for a given pure state $|\Phi\kt$
of the full system,
\begin{equation}
\rhoA=\tr_2\left(|\Phi\kt \br\Phi|\right), \;
\qquad
\rhoB=\tr_1\left(|\Phi\kt \br\Phi|\right),
\end{equation}
which are defined as the partial trace over the other subsystem.
For $N_1\leq N_2$ the reduced density matrices $\rhoA$ and $\rhoB$
have $N_1$ common eigenvalues $\lambda_j$,
which are called Schmidt eigenvalues
and obey the normalization condition
\begin{equation}
  \sum_{j=1}^{N_1} \lambda_j = 1.
\end{equation}
The remaining $N_2-N_1$ eigenvalues of $\rhoB$ are zero.
In the following we assume the Schmidt eigenvalues
to be ordered by decreasing value, i.e.\
$\lambda_1 \ge \cdots \ge \lambda_{N_1}$.
A state is unentangled if and only if $\lambda_1=1$
and all other Schmidt eigenvalues vanish.
If $\lambda_1<1$, the state is entangled, as it
is no longer represented as a product state.
Maximal entanglement is obtained when
$\lambda_j=1/N_1$ for all $j=1, ..., N_1$.

Based on the Schmidt eigenvalues one can define the moments
\begin{equation}
\label{eq:momentsdefn}
  \mu_{\alpha}= \sum_{j=1}^{N_1} \lambda_j^{\alpha}, \;\; \alpha >0,
\end{equation}
and the Havrda-Charv{\'a}t-Tsallis (HCT)
entropies~\cite{HavCha1967,Tsa1988,BenZyc2006} by
\begin{equation}
  S_{\alpha} = \dfrac{1-\mu_{\alpha}}{\alpha-1}.
\label{eq:Tsallis}
\end{equation}
In the limit of $\alpha\to 1$ the von Neumann entropy $S_1$
is obtained,
\begin{equation}
\begin{split}
  S_1 & = -\tr\left (\rhoA \ln \rhoA \right)
        = -\tr\left (\rhoB \ln \rhoB \right)\\
      & =- \sum_{j=1}^{N_1} \lambda_j \ln \lambda_j.
\end{split}
\end{equation}
States which are unentangled lead to $S_\alpha=0$
while a maximally entangled state for example leads to $S_1 = \ln N_1$.

For states chosen at random uniformly with
respect to the Haar measure from the full Hilbert space,
the average von Neumann entropy can be computed exactly
\cite{Pag1993, FooKan1994, San1995, Sen1996} and
has the large $N_1$ asymptotics
\begin{equation} \label{eq:Haar-S1-entanglement}
  \overline{S_1} = \ln N_1 -\frac{1}{2 N_2/N_1} .
\end{equation}
For the linear entropy $S_2$ the
exact finite--$N_1$ result is \cite{Lub1978}
\begin{equation} \label{eq:Haar-S2-entanglement}
  \overline{S_2} =  1-\frac{N_1 + N_2}{1+ N_1 N_2} .
\end{equation}

\subsection{Perturbative behavior of Schmidt eigenvalues}
           \label{subsec:schmidt_eigenvalues}

As for the uncoupled bipartite system, Eq.~\eqref{eq:U_bipartite_system}
for $\varepsilon=0$,
the states are not entangled, we have $\lambda_1=1$
and $\lambda_j=0$ for $j>1$.
For non-vanishing coupling, the states
become entangled such that $\lambda_1 < 1$
and the second largest Schmidt eigenvalue $\lambda_2$
gives the most relevant contribution.
To arrive at a general expression for the first two averaged
Schmidt eigenvalues we now follow
Refs.~\cite{LakSriKetBaeTom2016,TomLakSriBae2018}.
Based on Rayleigh-Schr\"odinger perturbation theory it has been shown
that the first two Schmidt eigenvalues can be approximated as
\begin{align}
  \lambda_1^{jk} = & \; 1 - \varepsilon^2 \sum_{j^{\prime}k^{\prime}\neq jk}
    \frac{|\bra{jk}V_{12}\ket{j^{\prime}k^{\prime}}|^2}
         {(\theta_{jk}-\theta_{j^{\prime}k^{\prime}})^2}
    \label{eq:lambda_1_perturbation}, \\
  \lambda_2^{jk} = & \; \varepsilon^2
  \frac{|\bra{jk}V_{12}\ket{\jkCN}|^2}{(\theta_{jk}-\theta_{\jkCN})^2} .
  \label{eq:lambda_2_perturbation}
\end{align}
The notation is the same as introduced in
Sec.~\ref{subsec:level_spacing_stat}.
Furthermore $\ket{\jkCN}$ is the state for which $\theta_{\jkCN}$ is closest
to $\theta_{jk}$.
Again $|\bra{jk}V_{12}\ket{j^{\prime}k^{\prime}}|^2$
is replaced by $\tilde{v}^2 \omega_{j^{\prime}k^{\prime}}$.
In addition
$(\theta_{jk}-\theta_{j^{\prime}k^{\prime}})/D$
is  substituted by
$\sRTwo_{j^{\prime}k^{\prime}}$
and $(\theta_{jk}-\theta_{\jkCN})/D$ by the closer neighbor
spacing $\sCN_{\jkCN}$.
Using $v^2 = \varepsilon^2 \tilde{v}^2$ leads to
\begin{align}
  \lambda_1^{jk} &\approx  \;  1 - \Lambda \sum_{j^{\prime}k^{\prime}\neq jk}
  \frac{\omega_{j^{\prime}k^{\prime}}}{\left(\sRTwo_{j^{\prime}k^{\prime}}\right)^2},\\
  \lambda_2^{jk} &\approx  \;
   \Lambda \; \frac{\omega_{\jkCN}}{\left(\sCN_{\jkCN}\right)^2}\; .
\end{align}
In order to find the perturbative behavior of the
first two Schmidt eigenvalues, one averages over
the random variables $\omega_{j^{\prime}k^{\prime}}$ and $\sRTwo_{j^{\prime}k^{\prime}}$
respectively $\sCN_{\jkCN}$.
The distribution of $\omega_{j^{\prime}k^{\prime}}$
is given by Eq.~\eqref{eq:RMTE-KT-coupling-K0-distrib}.
In order to perform the averaging for the first Schmidt eigenvalue define
\begin{align}
  R_{jk}(s, \omega) &= \sum_{j^{\prime}k^{\prime}\neq jk} \delta(\omega -
    \omega_{j^{\prime}k^{\prime}})\delta(s - \sRTwo_{j^{\prime}k^{\prime}})
\end{align}
as probability density to find a level with distance
$\sRTwo_{j^{\prime}k^{\prime}}$ to $\theta_{jk}$ and a corresponding
matrix element $\omega$ with the value $\omega_{jk}$.
Under the assumption that the matrix elements and the spacings
are uncorrelated, the ensemble average gives
\begin{align}
  \overline{R(s, \omega)} = \RhoOmega(\omega) R_2(s) = \RhoOmega(\omega) .
\end{align}
In the last equality the result
\begin{equation}
  R_2(s) = \overline{\sum_{j^{\prime}k^{\prime}\neq jk} \delta(s -
                \sRTwo_{j^{\prime}k^{\prime}})}
        =1
  \nonumber
\end{equation}
for a Poisson distributed random variable with $-\infty \leq s \leq \infty$
is used.
The distribution of the closer neighbor level spacings
$\sCN_{\jkCN}$ is given by \cite{SriLakTomBae2019}
\begin{equation} \label{eq:P-s-CN}
  \rho_{\text{CN}}(s) = 2 \text{exp}(-2s)\;.
\end{equation}
Thus the result for the averaged Schmidt eigenvalues is
\begin{align}
  \overline{\lambda_1}
    &= 1 - \Lambda \int_{-\infty}^{\infty} \ud s\int_0^{\infty}\ud \omega
       \; \frac{\omega}{{s}^2} \RhoOmega(\omega)
       \label{eq:lambda_1_perturbation_int},\\
  \overline{\lambda_2}
    &= \Lambda \int_0^{\infty} \ud s \int_0^{\infty} \ud \omega
       \; \frac{\omega}{{s}^2} \RhoOmega(\omega) \RhoCN(s)\;.
       \label{eq:lambda_2_perturbation_int}
\end{align}
Due to the singularity of the integrands for $s \rightarrow 0$, the
integrals in Eq.~\eqref{eq:lambda_1_perturbation_int} and
Eq.~\eqref{eq:lambda_2_perturbation_int} diverge.
Following Ref.~\cite{TomLakSriBae2018} we therefore perform the replacement
\begin{equation}
  \frac{\Lambda \omega}{s^2} \rightarrow \frac{1}{2} \Biggl( 1 -
  \frac{|s|}{\sqrt{s^2 + 4\Lambda \omega}} \Biggl) .
\end{equation}
This regularization is the correct description for small $s$ and has
the same asymptotics for large $s$. Therefore
this replacement is used in the entire domain of integration.

\begin{figure}[t]
  \includegraphics{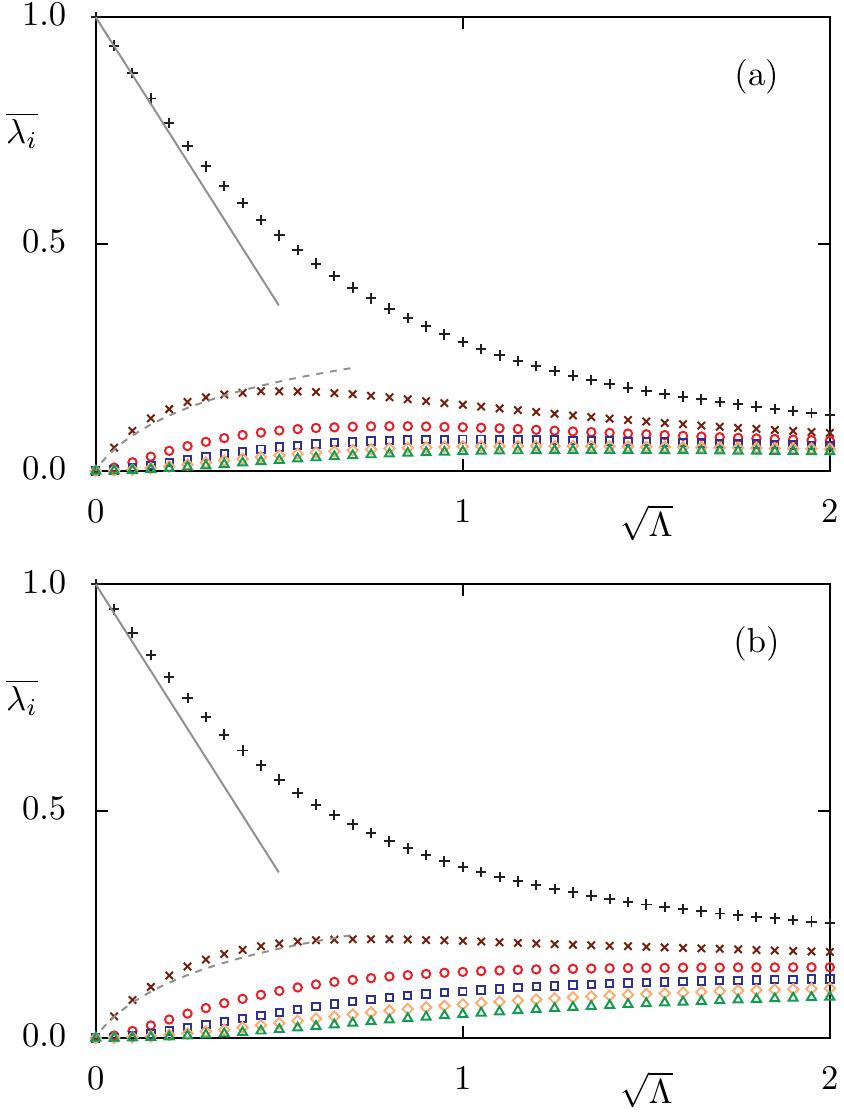}
   \caption{Average Schmidt eigenvalues $\overline{\lambda_i}$ in
            dependence on $\sqrt{\Lambda}$ for $i=1, 2, ..., 6$
            (top to bottom with different symbols)
            for the coupled kicked tops for
            (a) equal dimensions $j_1=j_2=50$ and
            (b) different dimensions$(j_1, j_2) = (3, 700)$.
            The solid grey line shows the
            prediction~\eqref{eq:pred_lambda_1}
            for $\overline{\lambda_1}$,
            the dashed grey line shows the prediction~\eqref{eq:pred_lambda_2}.
            Parameters are
            $k_1=12.0$, $k_2=15.0$, $\alpha_1 = 0.35$, and
            $\alpha_2 = 0.4$.
      }
    \label{fig:lambda_i}
\end{figure}

For the coupling \eqref{eq:U_12_kickedtop_4D}
for the coupled kicked tops
the distribution $\RhoOmega(\omega)$
is that of the random matrix transition ensemble
with product phases defined in Sec.~\ref{subsec:RMTE-KT-coupling}
and therefore given by Eq.~\eqref{eq:RMTE-KT-coupling-K0-distrib}.
Using this in Eqs.~\eqref{eq:lambda_1_perturbation_int}
and \eqref{eq:lambda_2_perturbation_int},
together with Eq.~\eqref{eq:P-s-CN}, we obtain
\begin{align}
  \overline{\lambda_1} &= 1 - \int_0^{\infty} \ud s \int_0^{\infty} \ud \omega
    \left( 1 - \frac{s}{\sqrt{{s}^2 + 4\Lambda\omega}}\right)
    \frac{\text{K}_0(\sqrt{\omega})}{\pi \sqrt{\omega}} \nonumber \\
            &= 1 - \frac{4}{\pi} \sqrt{\Lambda},
               \label{eq:pred_lambda_1}\\
\intertext{and} 
  \overline{\lambda_2} &= \int_0^{\infty} \ud s \int_0^{\infty} \ud \omega
            \left( 1 - \frac{s}{\sqrt{{s}^2 + 4\Lambda\omega}}\right) \ue^{-2s}
            \frac{\text{K}_0(\sqrt{\omega})}{\pi \sqrt{\omega}} \;.
            \label{eq:pred_lambda_2}
\end{align}
A comparison of these predictions with the average Schmidt eigenvalues
$\overline{\lambda_i}$, where the average is done over all eigenstates,
of the coupled kicked tops is shown in Fig.~\ref{fig:lambda_i}(a).
Here, we evaluate the integral in Eq.~\eqref{eq:pred_lambda_2} numerically.
Good agreement for small values of $\sqrt{\Lambda}$ is found.

Note that the predictions \eqref{eq:pred_lambda_1} and \eqref{eq:pred_lambda_2}
are based on the specific coupling \eqref{eq:rmt-TE-KT-diagonal-coupling}.
If one uses the coupling \eqref{eq:rmt-TE-diagonal-coupling}
introduced for the random matrix transition ensemble
in Ref.~\cite{SriTomLakKetBae2016}, see Sec.~\ref{subsec:RMTE},
one obtains for the COE case the results given in
App.~\ref{app:lambda-i-for-COE},
while the results for the CUE were obtained in
Refs.~\cite{SriTomLakKetBae2016,TomLakSriBae2018}.

The above derivation equally applies
to the case of different dimensionalities.
Thus in Eqs.~\eqref{eq:pred_lambda_1} and \eqref{eq:pred_lambda_2}
only the correct transition parameter $\Lambda$, computed
via Eq.~\eqref{eq:lambda_coe} and
Eqs.~\eqref{eq:U_12_trace_kickedtop_approx}--\eqref{eq:U_12_norm_partial_trace_2_kickedtop_approx}, has to be used.
Already starting from $j_1=3$ and large $j_2$
good agreement is found, see Fig.~\ref{fig:lambda_i}(b),
though in comparison with Fig.~\ref{fig:lambda_i}(a) the regime of agreement
for $\overline{\lambda_1}$ is smaller.

\subsection{Entanglement entropies}\label{sec:entanglement_entropies}
\label{subsec:entropies}

First we consider the perturbative description
of the entanglement entropies $S_\alpha$ for small $\sqrt{\Lambda}$.
To use Eq.~\eqref{eq:Tsallis} an expression for the moments
$\mu_{\alpha}$ is required.
For this we split the sum in the definition
of $\mu_{\alpha}$ in Eq.~\eqref{eq:momentsdefn} into two parts and consider
$\overline{\lambda_1^{\alpha}}$ and $\sum_{j>1}\overline{\lambda_j^{\alpha}}$
separately.  For $\overline{\lambda_1^{\alpha}}$ it is shown in
Ref.~\cite{TomLakSriBae2018} that the leading order result can be written as
\begin{align}
  \overline{\lambda_1^{\alpha}}
   =&\; 1 +2 \int_{0}^{\infty} \ud s\int_0^{\infty}\ud \omega
    \; \RhoOmega(\omega) \nonumber\\
    & \qquad \times \left[\left(1-\frac{1}{2}
                        \left(1-\frac{s}{\sqrt{s^2+4\Lambda\omega}}\right)
                   \right)^{\alpha}-1\right]\\
    &  + O(\Lambda), \nonumber
\end{align}
and the corrections of order $O(\Lambda)$ are given as
\begin{align}
   & 2 \int_{0}^{\infty} \ud s_1 \int_0^{\infty}\ud \omega_1
    \int_{0}^{\infty} \ud s_2\int_0^{\infty}\ud \omega_2
         \RhoOmega(\omega_1)\RhoOmega(\omega_2)
    \nonumber\\
    & \times \Biggl[1+
    \left(1-\frac{f(s_1,\omega_1)}{2}-\frac{f(s_2,\omega_2)}{2}\right)^{\alpha}
    \nonumber\\
    &\;\;\;
      - \left(1-\frac{f(s_1,\omega_1)}{2}\right)^{\alpha}
    - \left(1-\frac{f(s_2,\omega_2)}{2}\right)^{\alpha}\Biggl]\;.
\end{align}
Here the abbreviation $f(s,\omega) = 1 - s/(\sqrt{s^2+4\Lambda\omega})$ is used.
Using the density $\RhoOmega(\omega)$
from Eq.~\eqref{eq:RMTE-KT-coupling-K0-distrib}
for the coupled kicked tops this leads to
\begin{align} \label{eq:lambda-1-alpha}
  \overline{\lambda_1^{\alpha}}
     = 1 - C_1(\alpha)\sqrt{\Lambda} + C_3(\alpha)\Lambda
\end{align}
with
\begin{align}
  C_1(\alpha) =&\; \frac{2}{\pi} \int_0^{\frac{1}{2}} \ud t \frac{1-
                  (1-t)^{\alpha}}{t^{3/2}(1-t)^{3/2}} \nonumber \\
              =&\; \frac{4\sqrt{2}}{\pi}\;{}_2F_1
                  \left(-\frac{1}{2},\frac{3}{2}-\alpha;
                        \frac{1}{2}; \frac{1}{2}\right) \;, \\
  C_3(\alpha) =&\; \frac{2}{\pi^2} \int_0^{\frac{1}{2}} \ud t_1
                  \int_0^{\frac{1}{2}} \ud t_2\nonumber\\
                 & \times \frac{1 + (1-t_1-t_2)^{\alpha}
                                -(1-t_1)^{\alpha}-(1-t_2)^{\alpha}}
                  {t_1^{3/2}(1-t_1)^{3/2}t_2^{3/2}(1-t_2)^{3/2}}\;.
\end{align}
Here ${}_2F_1$ is Gauss' hypergeometric function
\cite[Eq.~15.2.1]{DLMFCurrent}.
For $\alpha=1$ Eq.~\eqref{eq:lambda-1-alpha} reproduces the prediction
Eq.~\eqref{eq:pred_lambda_1} for $ \overline{\lambda_1}$.
To calculate $\sum_{j>1}\overline{\lambda_j^{\alpha}}$
we use \cite{TomLakSriBae2018}
\begin{equation}
  \sum_{j>1}\overline{\lambda_j^{\alpha}} = \int_{-\infty}^{\infty}\ud s
  \int_0^{\infty}\ud \omega \;\frac{\RhoOmega(\omega)}{2^{\alpha}}
  \left( 1 - \frac{|s|}{\sqrt{s^2 + 4\Lambda \omega}}\right)^{\alpha} \;.
\end{equation}
Inserting Eq.~\eqref{eq:RMTE-KT-coupling-K0-distrib} for $\RhoOmega(\omega)$
leads to
\begin{align}
  \sum_{j>1}\overline{\lambda_j^{\alpha}}
       &=C_2(\alpha)\sqrt{\Lambda}\;\\
\intertext{with}
  C_2 (\alpha)
       &= \frac{2}{\pi}\int_0^{\frac{1}{2}} \frac{t^{\alpha}}{t^{3/2}(1-t)^{3/2}}
                  \nonumber\\
       &=  \frac{2}{\pi} B_{1/2}(\alpha-\frac{1}{2},-\frac{1}{2})\;.
\end{align}
Here $B_z(a, b)$ is the incomplete Beta function
\cite[Eq.~8.17.1]{DLMFCurrent}.
With this it is now possible to write the average moments
$\overline{\mu_{\alpha}}$ as
\begin{align}\label{eq:moments_pertubation}
  \overline{\mu_{\alpha}} &= 1- C(\alpha) \sqrt{\Lambda} + C_3(\alpha) \Lambda\;,
\end{align}
where
\begin{align} \label{eq:C-alpha}
  C(\alpha) &= C_1(\alpha)-C_2(\alpha)
             = \frac{4}{\sqrt{\pi}} \frac{\Gamma(\alpha -\frac{1}{2})}
                                         {\Gamma(\alpha-1)}\;.
\end{align}
This results in
\begin{equation}
  \overline{S_{\alpha}} = \frac{4}{\sqrt{\pi}}
    \frac{\Gamma(\alpha-\frac{1}{2})}{\Gamma (\alpha)} \sqrt{\Lambda}
    - \frac{C_3(\alpha)}{\alpha-1} \Lambda
    \label{eq:Tsallis_perturbation}
\end{equation}
as an approximation of the entropies for small $\Lambda$.
An important special case is the von Neumann entropy
obtained in the limit $\alpha \rightarrow 1$, which gives
\begin{equation}
  \overline{S_1} = 4\sqrt{\Lambda} - \left(\frac{4}{\pi}-1\right)\Lambda\;.
\end{equation}

In addition to this perturbative description of $\overline{S_{\alpha}}$,
valid for small $\sqrt{\Lambda}$,
the recursively embedded perturbation theory can be applied following
Ref.~\cite{TomLakSriBae2018} to obtain a complete description of the
entropies as a function of $\Lambda$.
The underlying idea is that with increasing
$\Lambda$ successively more and more Schmidt eigenvalues become
relevant. This can be accounted for by a recursive
description which can be approximated by a differential
equation. Furthermore, the maximal values
of the entropies for the fully entangled situation are used,
which follow from the moments of the \Marcenko-Pastur
distribution of the Schmidt eigenvalues \cite{SomZyc2004}.
Restricting to $N = N_1 = N_2$ one has
\begin{align}
  \overline{S_1^{\infty}} &= \ln N - \frac{1}{2},
    \label{eq:max-entropies-S-1}\\
  \overline{S_{\alpha}^{\infty}}
        &=\frac{1-\mathcal{C}_{\alpha} N^{1-\alpha}}{\alpha - 1} \qquad
    \text{for}\; \alpha>1 ,
    \label{eq:max-entropies-S-alpha}
\end{align}
where $\mathcal{C}_{\alpha} = \frac{1}{\alpha+1} \binom{2\alpha}{\alpha}$
are Catalan numbers \cite[\S26.5.]{DLMFCurrent}.
This leads to
\begin{align} \label{eq:S-alpha-recursive}
  \overline{S_{\alpha}(\Lambda)} \approx
    \left[1-\exp\left(-\frac{C(\alpha)}{(\alpha-1)\overline{S_{\alpha}^{\infty}}}
    \sqrt{\Lambda}\right)\right]\overline{S_{\alpha}^{\infty}}
\end{align}
as prediction for the entropies. In particular,
using $\lim_{\alpha\rightarrow1}\frac{C(\alpha)}{\alpha-1} = 4$
gives for the von Neumann entropy,
\begin{align} \label{eq:S-1-recursive}
  \overline{S_{1}(\Lambda)} \approx
    \left[1-\exp\left(-\frac{4}{\overline{S_{1}^{\infty}}}
    \sqrt{\Lambda}\right)\right]\overline{S_{1}^{\infty}}.
\end{align}

\begin{figure}[t]
  \includegraphics{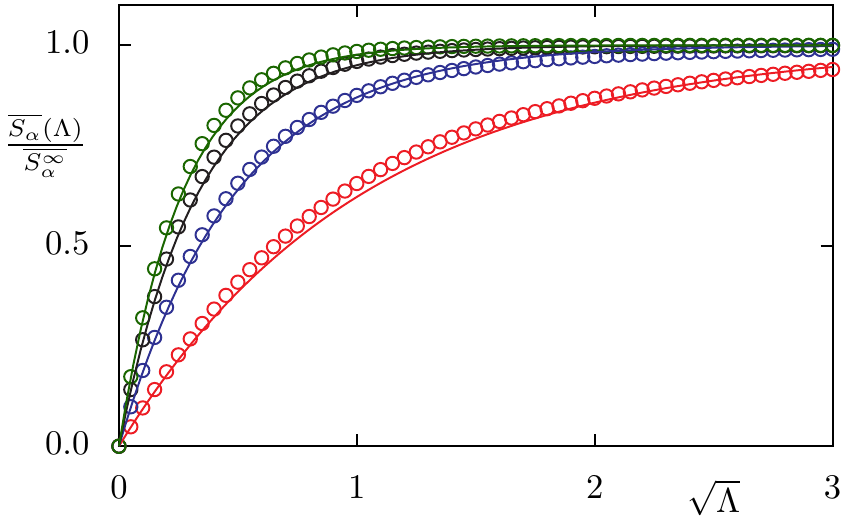}
  \caption{Average entanglement entropies $\overline{S_{\alpha}}$,
            rescaled by $\overline{S_{\alpha}^\infty}$,
            in dependence on $\sqrt{\Lambda}$ for $\alpha=1, 2, 3, 4$
            (bottom to top)
            for the coupled kicked tops.
            The solid curves show the prediction
            as given by Eq.~\eqref{eq:S-alpha-recursive} and
            Eq.~\eqref{eq:S-1-recursive}.
            Parameters are $j_1=j_2=50$,
            $k_1=12.0$, $k_2=15.0$, $\alpha_1 = 0.35$, and
            $\alpha_2 = 0.4$.
      }
    \label{fig:entropies}
\end{figure}

Figure~\ref{fig:entropies} shows a comparison of the recursively embedded
perturbation theory predictions with the results for the coupled kicked
tops. The agreement between the curves is overall very good.

\subsection{Entanglement entropies for different dimensions}
\label{subsec:entropies_diff_dim}

To describe the entanglement
entropies for different subsystem dimensions the recursively
embedded perturbation theory can be applied as well.
We restrict to the case of the linear entropy obtained for $\alpha=2$.
For this the maximum of the entropy $\overline{S_2^\infty}$ is exactly given by
Lubkin's result \eqref{eq:Haar-S2-entanglement}.
This leads to the prediction for the linear entropy
\begin{equation}
  \overline{S_{2}(\Lambda)}
  \approx
    \left[1-\exp\left(- \frac{2}{\overline{S_{2}^{\infty}}} \sqrt{\Lambda}\right)
    \right] \overline{S^{\infty}_{2}} ,
  \label{eq:S_2_diff_dim_scaled}
\end{equation}
as by Eq.~\eqref{eq:C-alpha}
one has $C(2) = 2$.
Thus while the functional dependence is the same
as in Eq.~\eqref{eq:S-alpha-recursive},
the different dimensionalities of the subsystems
are accounted for by the formula for $\overline{S_{2}^{\infty}}$
and the dependence of the transition parameter $\Lambda$
on the subsystem dimensions.
Note that for the other entropies with $\alpha \neq 2$,
the maximal values of the entropies
corresponding to Eqs.~\eqref{eq:max-entropies-S-1}
and \eqref{eq:max-entropies-S-alpha}
follow from the results in Ref.~\cite{SomZyc2004}.

\begin{figure}
  \includegraphics{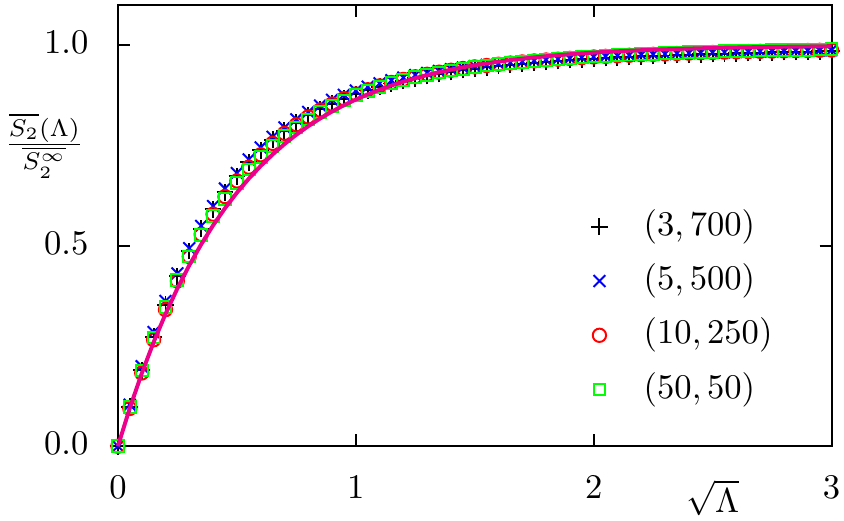}
  \caption{Rescaled linear entropy, $\overline{S_2} / \overline{S_2^{\infty}}$,
           in dependence on $\sqrt{\Lambda}$ for different
            dimensions $(j_1, j_2)=
            (3, 700),
            (5, 500),
            (10, 250),
            (50, 50)$
            of the coupled kicked tops.
            The solid magenta curve shows
            Eq.~\eqref{eq:S_2_diff_dim_scaled}
            using $\overline{S_2^\infty} = 1$ which corresponds
            to $N_1, N_2\to \infty$.
            Parameters are
            $k_1=12.0$, $k_2=15.0$, $\alpha_1 = 0.35$, and
            $\alpha_2 = 0.4$.
      }
    \label{fig:entropies-diff-dim}
\end{figure}

Figure~\ref{fig:entropies-diff-dim}
shows the rescaled linear entropy for the coupled kicked tops
for several pairs of different dimensions
as well as the prediction
from Eq.~\eqref{eq:S_2_diff_dim_scaled}.
From this plot it can be seen that for $j_1 \ge 3$
the linear entropy of the coupled
kicked tops for different dimensions, after rescaling by the corresponding
$\overline{S^{\infty}_{2}}$  given by Eq.~\eqref{eq:Haar-S2-entanglement},
collapse rather well to one universal curve
described by Eq.~\eqref{eq:S_2_diff_dim_scaled}.

Thus we get a remarkable range of universal behavior
and agreement with the theory \eqref{eq:S_2_diff_dim_scaled}.
Only for the very small system sizes $j_1=1, 2$ (not shown)
there are systematic differences
and a detailed understanding and theoretical
description in this case
is an interesting open question for the future.

\subsection{Statistics of Schmidt eigenvalues}
\label{subsec:stat_schmidt_eigenvalues}

The average Schmidt eigenvalues and the average
entanglement entropies provide a compact characterization
of the possible amount of entanglement
in dependence of the universal scaling parameter $\Lambda$.
More detailed information is obtained by considering
the statistics of the whole spectrum of Schmidt eigenvalues
\cite{BanLak2002}, or the entanglement spectrum \cite{LiHal2008}.
For large $\Lambda$ one expects that the distribution of the
scaled  Schmidt eigenvalues
\begin{equation} \label{eq:re-scaled-Schmidt-evals}
  x_i=\lambda_i \, N_1
\end{equation}
is given by the \Marcenko-Pastur distribution,
when $N_1$ and $N_2$ are large but their ratio $Q=N_2/N_1 \ge 1$ is fixed
\cite{SomZyc2004}.
This distribution reads
\cite{MarPas1967}
\begin{equation}
 \label{eq:marcenko-pastur-law}
P_{\text{MP}}^{Q}(x)
  = \frac{Q}{2\pi} \frac{\sqrt{(x_{+}-x)(x-x_{-})}}{x},
     \; x_{-} \le x \le x_{+},
\end{equation} where
\begin{equation}
x_{\pm}=1+\frac{1}{Q} \pm \frac{2}{\sqrt{Q}} .
\end{equation}
For chaotic states of coupled kicked tops, i.e.\ in the regime
of large $\Lambda$,  this has
been verified in Ref.~\cite{BanLak2002}.
Exact results for finite $N_1$ were
obtained in Refs.~\cite{KubAdaTod2008,KubAdaTod2013}.

\begin{figure}
  \includegraphics{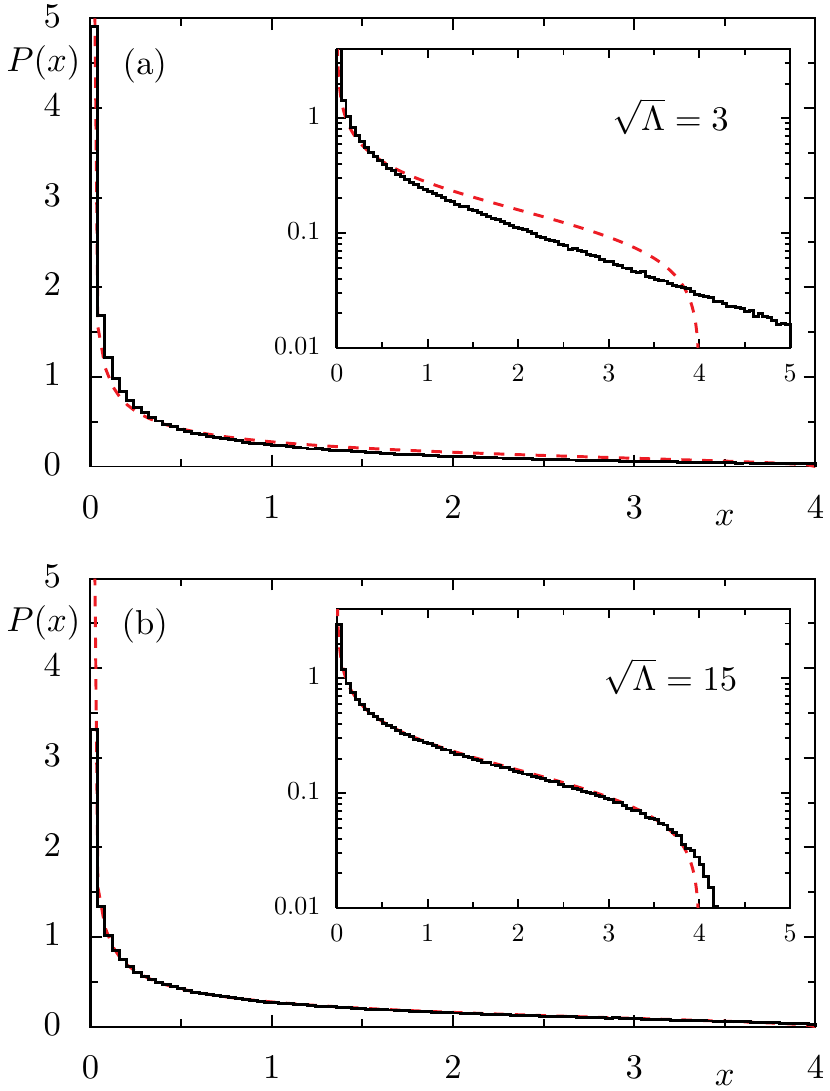}
   \caption{Distribution $P(x)$ of the re-scaled Schmidt eigenvalues
           $x_i = \lambda_i N_1$ for (a) $\sqrt{\Lambda}=3$
           and (b) $\sqrt{\Lambda}=15$, both with $N_1=2j_1+1$ for
           $j_1=j_2=50$.
           The red dashed curve shows the \Marcenko-Pastur distribution
           \eqref{eq:marcenko-pastur-law} for $Q=1$.
           The insets show the same data in a semi-logarithmic plot.
      }
    \label{fig:distrib-lambda-i}
\end{figure}

\begin{figure}
  \includegraphics{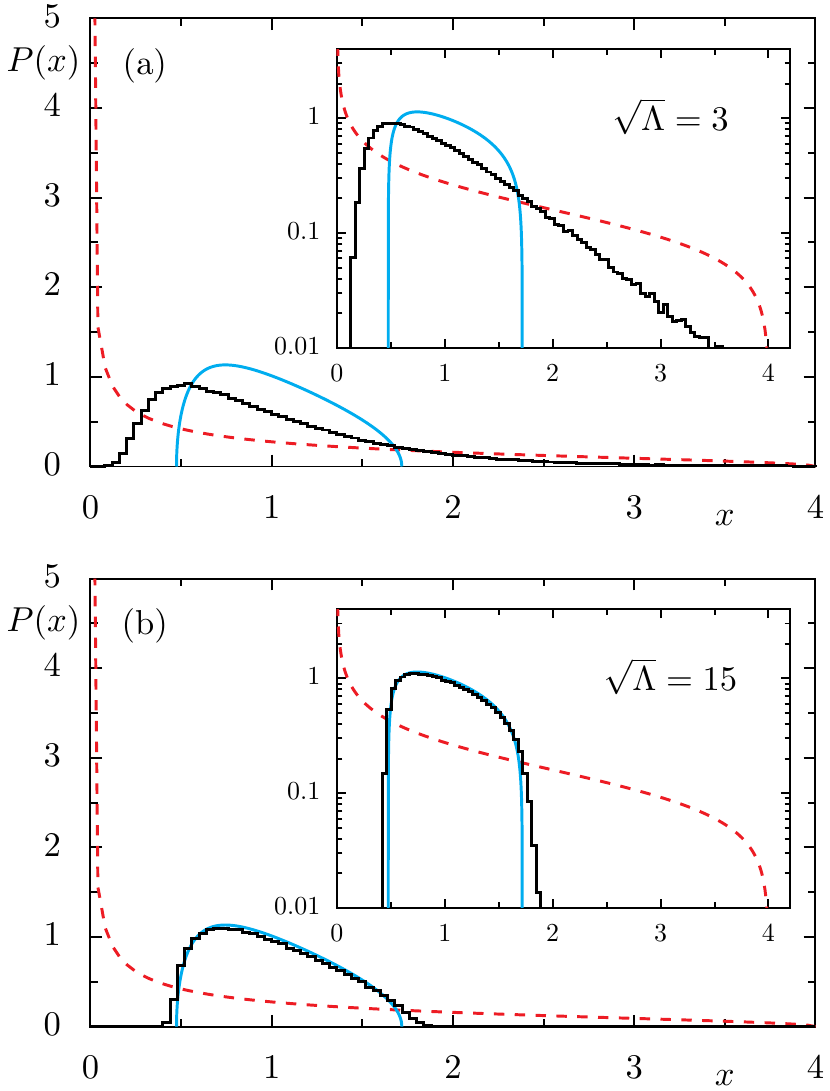}
   \caption{Distribution $P(x)$ of the re-scaled Schmidt eigenvalues
           $x_i = \lambda_i N_1$ for (a) $\sqrt{\Lambda}=3$
           and (b) $\sqrt{\Lambda}=15$, both with $N_1=2j_1+1$ for
           $(j_1, j_2) = (15, 160)$.
           The full cyan
           curve shows the \Marcenko-Pastur distribution
           \eqref{eq:marcenko-pastur-law}
           for $Q=(2j_1+1)/(2j_2+1) = 0.0966$
           and the red dashed curve for $Q=1$.
           The insets show the same data in a semi-logarithmic plot.
      }
    \label{fig:distrib-lambda-i-diff-dim}
\end{figure}

We want to investigate the dependence of the distribution of the
scaled Schmidt eigenvalues
on $\Lambda$ for equal subsystem dimensions as well as for different dimensions.
In the uncoupled case, i.e.\ at $\Lambda=0$,
all eigenstates are unentangled such that
$\lambda_1=1$ and $\lambda_i=0$ for $i>1$
leading to  $P(x) = (1 - 1/N_1) \delta(x) + 1/N_1 \delta (x-N_1)$.
With increasing $\sqrt{\Lambda}$ the distribution $P(x)$
of the re-scaled Schmidt eigenvalues \eqref{eq:re-scaled-Schmidt-evals}
will move towards
the \Marcenko-Pastur distribution.
Figure~\ref{fig:distrib-lambda-i} shows
the result for the case of equal dimension of the subsystems
and illustrates that this transition is
rather slow as even for $\sqrt{\Lambda}=15$
small deviations are visible near $x=4$.
These deviations are due to the finite system size
and become smaller with increasing $j_1=j_2$.
The slow convergence with $\sqrt{\Lambda}$
to the \Marcenko-Pastur distribution has already been observed for
the example of the coupled standard maps in Ref.~\cite{TomLakSriBae2018},
for which, however, the transition appears to be slightly
faster, which is consistent with the observations
for the spectral statistics made in Sec.~\ref{subsec:level_spacing_stat}.

The case of different dimensions of the subsystems is shown in
Fig.~\ref{fig:distrib-lambda-i-diff-dim}.
Again for rather large $\sqrt{\Lambda}$ good
agreement between the distribution for the coupled kicked tops and the
\Marcenko-Pastur distribution is found.
Interestingly, the distribution is quite concentrated around $x=1$,
so that one could think that there are
some states which are close to maximal entanglement,
i.e.\ $\lambda_i = 1/N_1$ for all $i$.
However, we observe, that this is not the case at least for a
finite $N_2$. To obtain maximal entanglement, sophisticated
protocols are needed \cite{BenBerPopSch1996}.

\section{Summary and outlook}\label{sec:summary}

For bipartite systems
the spectral statistics and entanglement of eigenstates
are investigated in dependence of a tunable interaction.
We focus on classically fully chaotic subsystem which can be modeled by
circular unitary or orthogonal ensembles and derive an exact expression
for the ensemble average of the transition parameter.
By specifying the statistical properties of the coupling between
the subsystems, different random matrix transition ensembles
are obtained.
In particular assuming a product structure for the coupling allows for
explicitly describing the dependence of the transition parameter on the
individual subsystems Hilbert space dimensions and the coupling strength.
An important model system following COE statistics is given by two coupled
kicked tops.
We utilize this system in order to illustrate the transition from
non-interacting to random matrix behavior.
To this end we consider the level spacing distribution in the case of equal
and unequal dimensions of the subsystems.
For equal dimensions the statistics depends solely on the transition parameter.
For unequal dimensions we find deviations if one
of the subsystems has a small dimension.
However, universality is already achieved when the smaller subsystem
has dimension larger than ten.
For large transition parameter the limiting case of Wigner distributed
level spacings is approached slower than for systems described by the CUE
transition ensemble.
A perturbative description, obtained from the
random matrix transition ensemble with product phases,
is in good agreement with numerical results.

For the average entanglement of eigenstates, in terms of their HCT entropies
including the von Neumann entropy, a universal scaling for both
equal and unequal dimensions is found.
Only if the dimension of one subsystem is smaller than five,
deviations from universality are observed.
Applying perturbation theory for the average first
and second Schmidt eigenvalues gives very
good agreement with the numerical results for the coupled kicked tops
for equal and unequal dimensions for small transition parameters.
Using the recursively embedded perturbation theory
allows to extend the perturbative description towards the large
coupling regime. Very good agreement of the HCT entropies and
the von Neumann entropy with the numerical results
is found for all values of the transition parameter.
Finally we study the distribution of Schmidt eigenvalues
for which a rather slow
transition from the unentangled case towards \Marcenko-Pastur
distribution for both equal and unequal dimensions is observed.

The results presented in this paper confirm that the
theory based on the transition parameter gives rise to an accurate
description of eigenstate entanglement for bipartite systems satisfying a
unitary symmetry when the specific structure of the coupling is taken
into account.
There are several interesting open questions.
The observed deviations from universality if one subsystem
is very small are not well captured by the asymptotic results
of the random matrix transition ensemble, but potentially
are accessible by analytic approaches.
Thus studying those small systems may give rise to further insight both in
terms of eigenstate entanglement as well as in the
time evolution of initially pure states.
Another interesting question for the future is to
find an analytical expression for the
transition between the distribution of the Schmidt eigenvalues
at $\Lambda=0$ and the \Marcenko-Pastur distribution
based on the random matrix transition ensemble.
Furthermore, as the system of two coupled kicked tops may be interpreted
as the collective dynamics of two spin chains with non-local
interaction one may ask to what extent the results
transfer to interacting many-body systems.

\acknowledgments

We thank Roland Ketzmerick, Arul Lakshminarayan,
Jan Schmidt, Shashi Srivastava, and Steve Tomsovic
for useful discussions.

\appendix

\section{Computation of the transition parameter}
\label{app:transition_parameter}

\subsection{COE} \label{app:COE-transition-parameter}

If the bipartite system is such that the individual
subsystems are described by random matrix theory
in the presence of an anti-unitary symmetry, one
can set up the random matrix transition ensemble
\eqref{eq:U_rmt-general} as
\begin{equation}
    U_{\text{COE}}(\varepsilon)
        = U_{12}(\varepsilon) (U_1^{\text{COE}}\otimes U_2^{\text{COE}}),
    \label{eq:U_rmt}
\end{equation}
where $U_1^{\text{COE}}$ und $U_2^{\text{COE}}$
are independently chosen COE random matrices
of dimension $N_1\times N_1 $ and $N_2\times N_2$, respectively.
The interaction $U_{12} \equiv U_{12}(\varepsilon)$ is a diagonal unitary matrix
of dimension $N_1N_2\times N_1N_2$.
For the moment we do not yet specify
the statistics of its entries.

We now  determine the transition parameter $\Lambda = v^2 / D^2$,
where $v^2$ is the mean square of the off-diagonal elements for
$U_{12}$ in the basis in which $U_1^{\text{COE}} \otimes U_2^{\text{COE}}$
is diagonal and $D$ is the mean level spacing.
For this consider $\Omega_i =  E_i^\dagger U_i^{\text{COE}} E_i$,
where $E_i$ is the matrix containing the eigenvectors
of $U_i^{\text{COE}}$ as columns.
Defining $\Upsilon = E_1 \otimes E_2$
we get the representation of $U_{12}$ in the requested basis,
$\Gamma= \Upsilon^\dagger U_{12}\Upsilon$,
where
\begin{equation}
  \gamma_{il} = (\Gamma)_{il}
     = \sum_{j,k=1}^{N_1N_2} \upsilon_{ki}^* (U_{12})_{kj} \upsilon_{jl}
\end{equation}
with $\upsilon_{jl} = (\Upsilon)_{jl}$.Thus
\begin{equation}     \label{eq:v_2_definition}
\begin{split}
    v^2 & = \frac{\sum_{i,l=1}^{N_1N_2}
                  |\gamma_{il}|^2 - \sum_{i=1}^{N_1N_2} |\gamma_{ii}|^2}
                 {(N_1N_2)^2 - N_1N_2}\\
     &= \frac{N_1 N_2- \sum_{i=1}^{N_1N_2} |\gamma_{ii}|^2}
     {N_1N_2(N_1N_2- 1)},
\end{split}
\end{equation}
where in the second equality the unitarity of $\Gamma$ has been used.
Next the sum over the diagonal elements is determined
\begin{align}
    \sum_{i=1}^{N_1N_2} |\gamma_{ii}|^2 =& \sum_{i=1}^{N_1N_2}
     \sum_{k=1}^{N_1N_2} \upsilon_{ki}^* (U_{12})_{kk} \upsilon_{ki}
     \sum_{l=1}^{N_1N_2} \upsilon_{li} (U_{12}^*)_{ll} \upsilon_{li}^* \nonumber\\
     =&
      \sum_{k,l=1}^{N_1N_2} (U_{12})_{kk} (U_{12}^*)_{ll}
      \sum_{i=1}^{N_1N_2} |\upsilon_{ik}|^2 |\upsilon_{il}|^2 \,.
      \label{eq:v_2_sum_over_diagnoal_elements}
  \end{align}
Next the product $|\upsilon_{ik}|^2 |\upsilon_{il}|^2 $
is replaced by its average $\overline{|\upsilon_{ik}|^2 |\upsilon_{il}|^2 }$
over the COE. By the definition of $\Upsilon$
and the independence of $U_1^{\text{COE}}$ and $U_2^{\text{COE}}$ one gets
\begin{equation}
  \overline{|\upsilon_{ik}|^2 |\upsilon_{il}|^2 }
    = \overline{|(E_1)_{i_1k_1}|^2 |(E_1)_{i_1l_1}|^2}\;
      \overline{|(E_2)_{i_2k_2}|^2 |(E_2)_{i_2l_2}|^2} \,,
      \label{eq:v_2_whole_average}
\end{equation}
where, due to the product structure, we identify $i \equiv (i_1,i_2)$,
$k \equiv (k_1,k_2)$, and $l \equiv (l_1,l_2)$.
For each term one has in case of the COE \cite{UllPor1963b}
\begin{equation}     \label{eq:v_2_coe_average}
    \overline{|(E_1)_{i_1k_1}|^2 |(E_1)_{i_1l_1}|^2}
    = \frac{2\delta_{k_1l_1}+1}{N_1 (N_1 +2)} \,.
\end{equation}
Thus we get for the average in Eq.~\eqref{eq:v_2_whole_average}
\begin{equation}
\begin{split}        \label{eq:v_2_calc_whole_average}
    \overline{|\upsilon_{ik}|^2 |\upsilon_{il}|^2 }
        & = \frac{2\delta_{k_1l_1}+1}{N_1 (N_1 +2)} \;
            \frac{2\delta_{k_2l_2}+1}{N_2 (N_2 +2)} \\
       & = \frac{4\delta_{k_1l_1}\delta_{k_2l_2} + 2 \delta_{k_1l_1}
                 + 2 \delta_{k_2l_2} + 1}
       {N_1 (N_1 +2)N_2 (N_2 +2)} \,.
\end{split}
\end{equation}
As the right-hand side is independent of $i$, one gets
\begin{equation} \label{eq:v_2_sum_over_diagonal_elements_with_average}
\begin{split}
    \sum_{i=1}^{N_1N_2} |\gamma_{ii}|^2
     = &\sum_{k_1,l_1 = 1}^{N_1} \sum_{k_2,l_2 = 1}^{N_2} \\
     & \qquad(  4\delta_{k_1l_1}\delta_{k_2l_2} +
       2 \delta_{k_1l_1} +
       2 \delta_{k_2l_2} +
       1) \\
   & \qquad \times
    \frac{
      \bra{k_1k_2}U_{12}\ket{k_1k_2} \bra{l_1l_2}U_{12}^*\ket{l_1l_2}}
      {(N_1 +2)(N_2 +2)}
        \,.
\end{split}
\end{equation}
Using the partial traces and the Hilbert-Schmidt norm we arrive at
\begin{equation}
  \sum_{i=1}^{N_1N_2} |\gamma_{ii}|^2
   = \frac{4N_1N_2 + 2 ||U_{12}^{(1)}||^2 + 2 ||U_{12}^{(2)}||^2
      + |\text{tr}(U_{12})|^2}
  {(N_1+2)(N_2+2)} \,.
  \label{eq:v_2_sum_over_diagonal_elements_with_traces}
\end{equation}
Insertion
in Eq.~\eqref{eq:v_2_definition}
gives the final result \eqref{eq:lambda_coe}
for the transition parameter in the COE case.

\subsection{CUE} \label{app:CUE-transition-parameter}

The derivation for the CUE follows the same
steps as for the COE.
The only difference is to replace relation
\eqref{eq:v_2_coe_average} by the CUE result \cite[Eq.~(10)]{PucMis2017},
\begin{equation}     \label{eq:v_2_cue_average}
    \overline{|(E_1)_{i_1k_1}|^2 |(E_1)_{i_1l_1}|^2}
    = \frac{\delta_{k_1l_1}+1}{N_1 (N_1 +1)} \,.
\end{equation}
With this the result \eqref{eq:lambda_cue} is obtained.

\subsection{Random matrix transition ensemble}
\label{app:COE-CUE-ensemble-transition-parameter}

For the random matrix transition ensemble
introduced in Ref.~\cite{SriTomLakKetBae2016}, see Sec.~\ref{subsec:RMTE},
the interaction is given by the diagonal matrix with
random phases.
In this case the partial traces and Hilbert-Schmidt norms have been
derived in Refs.~\cite{SriTomLakKetBae2016,TomLakSriBae2018} as
\begin{align}
    \overline{|\text{tr}(U_{12}^{\text{RMT}})|^2}
     = & \, N_1N_2 \Bigl( 1 + (N_1N_2 -1)
        \frac{\text{sin}^2(\pi \varepsilon)}{\pi^2 \varepsilon^2} \Bigl) \;, \\
    \overline{|| U_{12}^{\text{RMT}^{(1)}}||^2} \,
     = &\, N_1N_2 \Bigl( 1 + (N_2 -1)
         \frac{\text{sin}^2(\pi \varepsilon)}{\pi^2 \varepsilon^2} \Bigl),  \\
    \overline{|| U_{12}^{\text{RMT}^{(2)}}||^2}
     = &\, N_1N_2 \Bigl( 1 + (N_1 -1)
      \frac{\text{sin}^2(\pi \varepsilon)}{\pi^2 \varepsilon^2} \Bigl) \,.
\end{align}
Using these in
Eq.~\eqref{eq:lambda_coe} and Eq.~\eqref{eq:lambda_cue}, respectively,
gives the corresponding
results Eq.~\eqref{eq:lambda-COE-TP} for the COE and
Eq.~\eqref{eq:lambda-CUE-a-la-TP} for the CUE.

\subsection{Random matrix transition ensemble with product phases}
\label{app:COE-CUE-ensemble-transition-parameter-product-phases}

To calculate the transition parameter for the random matrix transition ensemble
with product phases, see Sec.~\ref{subsec:RMTE-KT-coupling},
we first consider the squared norm of the trace and the
Hilbert-Schmidt norm of the partial traces of the interaction
$U_{12}(\varepsilon)$ with the coupling in
Eq.~\eqref{eq:rmt-TE-KT-diagonal-coupling}.
To this end we collect the random variables into the vectors
$\xi = (\xi_j)_{j=1,\ldots,N_1}$ and $\tilde{\xi} =
(\tilde{\xi}_k)_{k=1,\ldots,N_2}$
where all components are i.i.d. distributed uniformly on $\left[-1/2, 1/2\right]$.
Averaging over the random variables in the coupling allows for writing
\begin{align}
\label{eq:rmt-TE-trace-squared}
  \overline{|\text{tr}(U_{12})|^2} =
      \sum_{j,j^{\prime}=1}^{N_1} \sum_{k,k^{\prime}=1}^{N_2}
  \hspace*{-0.5cm}\int\limits_{\hspace*{0.7cm}\left[-1/2, 1/2\right]^{N_1 + N_2}}
  \hspace*{-1cm}\ud\xi \ud \tilde{\xi}\;
  \ue^{2\pi \ui \varepsilon (\xi_j\tilde{\xi}_k - \xi_{j^{\prime}}\tilde{\xi}_{k^{\prime}})}\;.
\end{align}
For each term in this fourfold sum the integrand depends at most on
four of the integration variables, namely if $j \neq j^{\prime}$ and
$k \neq k^{\prime}$, and we can integrate over the remaining random
variables each giving a factor of one.
This gives
\begin{align}
\hspace*{-0.5cm}\int\displaylimits_{\hspace*{0.6cm}\left[-1/2,
	1/2\right]^{4}}\hspace*{-0.9cm}\ud\xi_j\ud\xi_{j^{\prime}}
\ud\tilde{\xi}_k\ud\tilde{\xi}_{k^{\prime}}
\;\ue^{2\pi \ui \varepsilon (\xi_j\tilde{\xi}_k -
	\xi_{j^{\prime}}\tilde{\xi}_{k^{\prime}})}
= \frac{4}{\varepsilon^2\pi^2} \text{Si}\left(\frac{\varepsilon\pi}{2}\right)^2,
\end{align}
where Si$(x)$ is the sine integral \cite[Eq.~6.2.9]{DLMFCurrent}.
Moreover it does not depend on the values of $j$, $j^{\prime}$, $k$ and
$k^{\prime}$ and there are $N_1N_2(N_1N_2 - N_1N_2 + 1)$
possible combinations of indices for this case.
Furthermore, there are $N_1N_2(N_2 - 1)$ cases
for which $j= j^{\prime}$ and $k \neq k^{\prime}$
and $N_2N_1(N_1 - 1)$ cases
for which $j \neq j^{\prime}$ and $k = k^{\prime}$
and for which the integrand in
Eq.~\eqref{eq:rmt-TE-trace-squared} depends on three integration
variables only.
Finally, there are $N_1N_2$ cases for which $j = j^{\prime}$ and
$k = k^{\prime}$ where the integrand depends on two integration variables only.
In all cases the corresponding integrals can be evaluated analytically.
Combining the integrals and taking the frequency of their appearance into
account gives
\begin{align}
  \overline{|\text{tr}(U_{12})|^2} =&
  \, N_1N_2 \biggl(1 + \frac{2}{\varepsilon^2 \pi^2}
      \Bigl[
      (N_1+N_2-2)\chi(\varepsilon\pi) \nonumber\\
      &+2(N_1-1)(N_2-1)\text{Si}\left(\frac{\varepsilon\pi}{2}\right)^2
      \Bigl]\biggl)\;,
      \label{eq:U_12_trace_RMTE_product}
\end{align}
where $\chi(x) = x\text{Si}(x) + \cos(x) -1$ has been used as abbreviation.
Using the same arguments one finds
\begin{align}
  \overline{||U_{12}^{(1)}||^2} =& \,N_1N_2\left(1+(N_2-1)
      \frac{2\,\chi(\varepsilon\pi)}{\varepsilon^2 \pi^2} \right)\;,
      \label{eq:U_12_norm_partial_trace_1_RMTE_product}\\
  \overline{||U_{12}^{(2)}||^2} =& \,N_1N_2\left(1+(N_1-1)
      \frac{2\,\chi(\varepsilon\pi)}{\varepsilon^2 \pi^2} \right)\;.
      \label{eq:U_12_norm_partial_trace_2_RMTE_product}
\end{align}
The transition parameter for the COE follows by inserting
Eqs.~\eqref{eq:U_12_trace_RMTE_product},
\eqref{eq:U_12_norm_partial_trace_1_RMTE_product},
and
\eqref{eq:U_12_norm_partial_trace_2_RMTE_product}
in Eq.~\eqref{eq:lambda_coe},
and for the CUE via Eq.~\eqref{eq:lambda_cue}.

\subsection{Coupled kicked tops} \label{app:KT-transition-parameter}

To determine the transition parameter for the coupled kicked tops, the
partial traces and Hilbert-Schmidt norm
have to be computed for the interaction $U_{12}(\varepsilon)$
as defined in Eq.~\eqref{eq:U_12_kickedtop_4D}.
The result is
\begin{align}
    |\text{tr}(U_{12})|^2 =& \sum_{m_1,s_1 =-j_1}^{j_1} \sum_{m_2,s_2 =-j_2}^{j_2}
      \cE(s_1, s_2)
      \cE(m_1, -m_2)
      \;, \label{eq:U_12_trace_kickedtop}\\
    ||U_{12}^{(1)}||^2 =& \sum_{s_1 =-j_1}^{j_1} \sum_{s_2,m_2 =-j_2}^{j_2}
      \cE(s_1, s_2)
      \cE(s_1, -m_2)
      , \label{eq:U_12_norm_partial_trace_1_kickedtop}\\
    ||U_{12}^{(2)}||^2 =& \sum_{s_1, m_1 =-j_1}^{j_1} \sum_{s_2 =-j_2}^{j_2}
      \cE(s_1, s_2)
      \cE(m_1, -s_2)
      \;, \label{eq:U_12_norm_partial_trace_2_kickedtop}
  \end{align}
where
$\cE(s, m) = \exp(-\ui \frac{\varepsilon}{\sqrt{j_1j_2}} s m)$
has been used as abbreviation.

For large $j_1, j_2$ the sums can be approximated by integrals
which can be evaluated exactly, giving
\begin{align}
    |\text{tr}(U_{12})|^2 \approx&\,  \biggl(4 \frac{\sqrt{j_1j_2}}{\varepsilon}
      \, \text{Si}(\varkappa/2) \biggl)^2 \;,
      \label{eq:U_12_trace_kickedtop_approx}\\
    ||U_{12}^{(1)}||^2 \approx
    & \,8 \frac{j_1j_2}{\varepsilon^2N_1} \chi(\varkappa),
      \label{eq:U_12_norm_partial_trace_1_kickedtop_approx}\\
    ||U_{12}^{(2)}||^2 \approx
    & \, 8 \frac{j_1j_2}{\varepsilon^2N_2}\chi(\varkappa),
 \label{eq:U_12_norm_partial_trace_2_kickedtop_approx}
\end{align}
where $\varkappa := \frac{\varepsilon N_1N_2}{2\sqrt{j_1j_2}}$.
The explicit expression of the transition parameter $\Lambda$ is then
obtained using Eq.~\eqref{eq:lambda_coe}.

For very large $\varepsilon$
the transition parameter $\Lambda$, determined
from Eq.~\eqref{eq:lambda_coe}
and Eqs.~\eqref{eq:U_12_trace_kickedtop_approx}--\eqref{eq:U_12_norm_partial_trace_2_kickedtop_approx}
saturates with value
\begin{equation}
  \Lambda_{\text{max}} =
  \frac{N_1^2 N_2^2  (N_1  N_2 + 2  (N_1 + N_2))}
       {4 \pi^2  (N_1 N_2 - 1)  (N_1 + 2)  (N_2 + 2)} .
\end{equation}
Thus for fixed $N_1$ and $N_2$ it is not possible
to obtain arbitrarily large $\Lambda$.

Note that for the interaction \eqref{eq:U_12_kickedtop_4D} of the coupled
kicked tops choosing the specific value
$\varepsilon=2 \pi$
gives $U_{12}(\varepsilon)=\mathrm{Id}$.
This operator does not create any interaction
between the two kicked tops.
This illustrates the limits of the applicability
of the transition parameter which has been obtained from
perturbation theory and therefore provides the correct
description for small values of $\varepsilon$ only.

\section{COE random matrix transition ensemble}
\label{app:lambda-i-for-COE}

In this appendix for completeness we derive
results for the entropies for the
COE random matrix transition ensemble defined in
Eq.~\eqref{eq:rmt-TE-diagonal-coupling}.
The results for the CUE case have been obtained in
Ref.~\cite{LakSriKetBaeTom2016,TomLakSriBae2018}.\\
The perturbative behavior of $\lambda_1$ and $\lambda_2$ can be determined
following the steps in Sec.~\ref{subsec:schmidt_eigenvalues} using the
coupling \eqref{eq:porter-thomas}. The result is
\begin{align}
  \overline{\lambda_1} &=\, 1 - \int_0^{\infty} \ud s \int_0^{\infty} \ud \omega
    \Bigl( 1 - \frac{s}{\sqrt{s^2+4\Lambda \omega}} \Bigl)
    \frac{\ue^{-\omega/2}}{\sqrt{2\pi \omega}}\nonumber\\
    &=1- \frac{4}{\sqrt{2\pi}}\sqrt{\Lambda}
            \label{eq:pred_lambda_1-COE-std-coupling}
\end{align}
and
\begin{align}
    \overline{\lambda_2} &= \int_0^{\infty}  \ud s \int_0^{\infty} \ud \omega
       \Bigl( 1 - \frac{s}{\sqrt{s^2+4\Lambda \omega}} \Bigl) \ue^{-2s}
       \frac{\ue^{-\omega/2}}{\sqrt{2\pi \omega}} \;.
            \label{eq:pred_lambda_2-COE-std-coupling}
\end{align}
The perturbative description of the
entanglement entropies $S_{\alpha}$ is given by
the averaged moments $\overline{\mu_{\alpha}}$
in Eq.~\eqref{eq:moments_pertubation}
with the terms
\begin{align} \label{eq:C-alpha-mod}
  C(\alpha) &= C_1(\alpha)-C_2(\alpha)
             = 2\sqrt{2} \;\frac{\Gamma(\alpha -\frac{1}{2})}
                                         {\Gamma(\alpha-1)}\;,
\end{align}
using
\begin{align}
  C_1(\alpha) =&\; \sqrt{\frac{2}{\pi}} \int_0^{\frac{1}{2}} \ud t \frac{1-
                  (1-t)^{\alpha}}{t^{3/2}(1-t)^{3/2}} \nonumber \\
              =&\; \frac{4}{\sqrt{\pi}}\;{}_2F_1
                  \left(-\frac{1}{2},\frac{3}{2}-\alpha;
                        \frac{1}{2}; \frac{1}{2}\right) \;, \\
  C_2 (\alpha) =& \sqrt{\frac{2}{\pi}}\int_0^{\frac{1}{2}}
                  \frac{t^{\alpha}}{t^{3/2}(1-t)^{3/2}}
                        \nonumber\\
              =&  \sqrt{\frac{2}{\pi}} B_{1/2}(\alpha-\frac{1}{2},-\frac{1}{2}),
\end{align}
and
\begin{align}
    C_3(\alpha) =&\; \frac{1}{\pi} \int_0^{\frac{1}{2}} \ud t_1
                  \int_0^{\frac{1}{2}} \ud t_2\nonumber\\
                 & \times \frac{1 + (1-t_1-t_2)^{\alpha}
                                -(1-t_1)^{\alpha}-(1-t_2)^{\alpha}}
                  {t_1^{3/2}(1-t_1)^{3/2}t_2^{3/2}(1-t_2)^{3/2}}\;,
\end{align}
which takes the coupling \eqref{eq:porter-thomas} into account.

\section{Matrix element distribution for product structure}
\label{ap:matrix_element_dist_KT}

\begin{figure}[b]
  \includegraphics{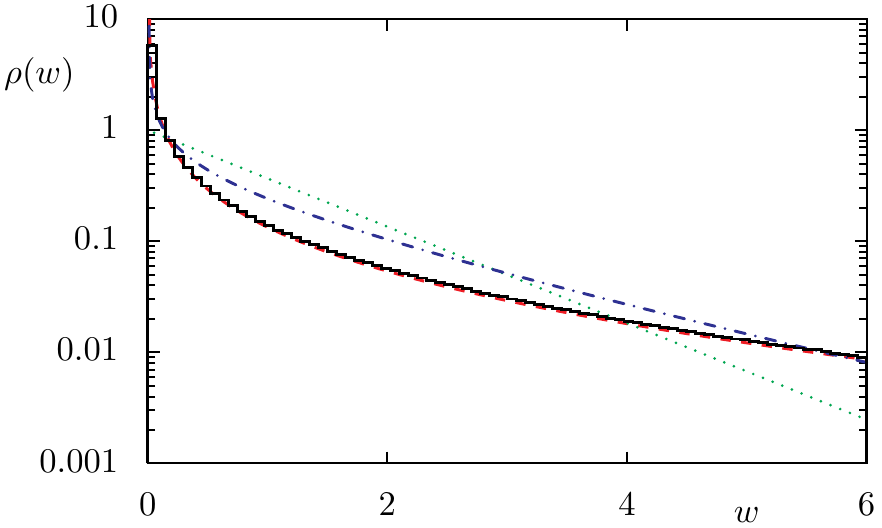}

  \caption{Matrix element distribution for the
  coupled kicked tops in comparison with the prediction
  \eqref{eq:matrix-elements-K0-prediction}, red dashed line,
  and  exponential distribution \eqref{eq:exponential},
  green dotted line,
  and the Porter-Thomas distribution \eqref{eq:porter-thomas},
  blue dash-dotted line.}
  \label{fig:matrix_ele_distr}
\end{figure}

In this appendix we derive the distribution of
the matrix elements \eqref{eq:omega-jk-def}
for the case that the coupling matrix $V_{12}$
has the product structure $V_{12} = V_1 V_2$
with $V_{\ell}$ only acting on the $\ell$-th subsystem.
This situation for example
occurs for the coupling \eqref{eq:rmt-TE-KT-diagonal-coupling}
of the random matrix transition ensemble with product phases
and the coupling \eqref{eq:U_12_kickedtop_4D}
of the coupled kicked tops.
As defined in Sec.~\ref{subsec:RMTE}, we have
\begin{align}
  \tilde{v}^2 \omega_{jk}
    = |\bra{jk}V_{12}\ket{j^{\prime} k^{\prime}}|^2
    = |\bra{jk}V_1 V_2\ket{j^{\prime} k^{\prime}}|^2\;,
       \nonumber
\end{align}
where in the second equality the product structure of $V_{12}$
has been used.
As $\ket{jk}$ and $\ket{j^{\prime} k^{\prime}}$ are eigenstates
of the uncoupled system, one can write
\begin{align}
  \tilde{v}^2 \omega_{jk}
    &= |\bra{jk}V_1 \ket{j^{\prime} k^{\prime}}|^2
       |\bra{jk}V_2 \ket{j^{\prime} k^{\prime}}|^2
       \nonumber\\
    &= \tilde{v_1}^2 \omega_{1_{jk}}\, \tilde{v_2}^2 \omega_{2_{jk}}
      = \tilde{v}^2 \omega_{1_{jk}}\omega_{2_{jk}}\;,\nonumber
\end{align}
which defines $\omega_{1_{jk}}$ and $\omega_{2_{jk}}$.
In the COE case both $\omega_{1_{jk}}$ and $\omega_{2_{jk}}$ follow
the Porter-Thomas distribution \eqref{eq:porter-thomas},
while for the CUE both obey the exponential \eqref{eq:exponential}.
Thus, for the distribution $\RhoOmega(\omega)$
of the matrix elements of $V_{12}$ one gets for the COE
\begin{align}
  \RhoOmega(\omega)
    &= \sum_{j^{\prime} k^{\prime} \neq jk} \delta(\omega-\omega_{j^{\prime} k^{\prime}})
       \nonumber \\
    &= \sum_{j^{\prime} k^{\prime} \neq jk} \delta(\omega -
       \omega_{1_{j^{\prime} k^{\prime}}} \omega_{2_{j^{\prime} k^{\prime}}})
        \nonumber\\
    &= \int \ud \omega_1 \ud \omega_2 \; \delta(\omega - \omega_1\omega_2)
        \frac{\ue^{-(\omega_1 + \omega_2)/2}}{2 \pi
        \sqrt{\omega_1 \omega_2}} \nonumber\\
    &= \int \ud \omega_2 \; \frac{\ue^{-(\frac{\omega}{\omega_2} + \omega_2)/2}}
       {2 \pi \omega_2 \sqrt{\omega}} \nonumber \\
    &= \frac{1}{\pi \sqrt{\omega}} K_0(\sqrt{\omega}) \;.
      \label{eq:matrix-elements-K0-prediction}
\end{align}
Here $K_\nu$ is the modified Bessel function of the second kind
\cite[Eq.~10.25.3]{DLMFCurrent}.
Similarly, one gets for the CUE case
\begin{align}
  \RhoOmega(\omega) = 2 K_0(2\sqrt{\omega})\;.
  \label{eq:matrix-elements-K0-prediction-CUE}
\end{align}

Figure~\ref{fig:matrix_ele_distr}
shows the matrix element distribution
for the coupled kicked tops in comparison
with the prediction \eqref{eq:matrix-elements-K0-prediction}.
Very good agreement is found, while
the random matrix transition ensemble \eqref{eq:U_rmt-general}
with coupling \eqref{eq:rmt-TE-diagonal-coupling}
gives in the CUE case the exponential distribution \eqref{eq:exponential}
and in the COE case the Porter-Thomas distribution \eqref{eq:porter-thomas}.

%

\end{document}